%% file: earth3.tex
\documentclass{aastex63}
\usepackage{amsmath}
\usepackage{amsfonts}
\usepackage{amssymb}
\usepackage{multirow}
\usepackage{graphicx}

\def \apjs {{\rm {ApJS}}}

\def \aap {{\rm {A\&A}}}

\def \apjl{\rm {ApJL}}

\def \deg{$^\circ$}

\newcommand{\ms}{\mbox{m s$^{-1}~$}}
\date{\today}
\received{xxx}
\revised{xxx}
\accepted{xxx}
\submitjournal{ApJS}

\shorttitle{Search for nearby Earth analogs}
\shortauthors{Feng et al.}

\begin{document}
\title{Search for Nearby Earth Analogs.\\II. detection of five
  new planets, eight planet candidates, and confirmation of three planets around nine nearby M dwarfs
  \footnote{This paper includes data gathered with the 6.5
    meter Magellan Telescopes located at the Las Campanas Observatory, Chile.}}
\author[0000-0001-6039-0555]{Fabo Feng}
\affiliation{Department of Terrestrial Magnetism, Carnegie Institution for Science, Washington, DC 20015, USA}
\author{R. Paul Butler}
\affiliation{Department of Terrestrial Magnetism, Carnegie Institution for Science, Washington, DC 20015, USA}
\author{Stephen A. Shectman}
\affiliation{Observatories of the Carnegie Institution for Science, 813 Santa Barbara St., Pasadena, CA 91101}
\author{Jeffrey D. Crane}
\affiliation{Observatories of the Carnegie Institution for Science, 813 Santa Barbara St., Pasadena, CA 91101}
\author{Steve Vogt}
\affiliation{UCO/Lick Observatory, University of California, Santa Cruz, CA 95064,USA}
\author{John Chambers}
\affiliation{Department of Terrestrial Magnetism, Carnegie Institution for Science, Washington, DC 20015, USA}
\author{Hugh R. A. Jones}
\affiliation{Centre for Astrophysics Research, University of Hertfordshire, College Lane, AL10 9AB, Hatfield, UK}
\author{Sharon Xuesong Wang}
\affiliation{Observatories of the Carnegie Institution for Science, 813 Santa Barbara St., Pasadena, CA 91101}
\author{Johanna K. Teske}
\affiliation{Department of Terrestrial Magnetism, Carnegie Institution for Science, Washington, DC 20015, USA}
\affiliation{Observatories of the Carnegie Institution for Science, 813 Santa Barbara St., Pasadena, CA 91101}
\affiliation{Hubble Fellow}
\author{Jenn Burt}
\affiliation{Department of Physics, and Kavli Institute for Astrophysics and Space Research, M.I.T., Cambridge, MA 02139, USA}
\author{Mat\'ias R. D\'iaz}
\affiliation{Observatories of the Carnegie Institution for Science, 813 Santa Barbara St., Pasadena, CA 91101}
\affiliation{Departamento de Astronom\'ia, Universidad de Chile, Camino El Observatorio 1515, Las Condes, Santiago, Chile}
\author{Ian B. Thompson}
\affiliation{Observatories of the Carnegie Institution for Science, 813 Santa Barbara St., Pasadena, CA 91101}

\correspondingauthor{Fabo Feng}
\email{ffeng@carnegiescience.edu}

\begin{abstract}
\cite{zechmeister09} surveyed 38 nearby M dwarfs from March
2000 to March 2007 with VLT2 and the UVES spectrometer.
This data has recently been reanalyzed \citep{butler19},
yielding a significant improvement in the Doppler velocity
precision.  Spurred by this, we have combined the UVES data
with velocity sets from HARPS, Magellan/PFS, and Keck/HIRES.
Sixteen planet candidates have been uncovered orbiting nine M
dwarfs. Five of them are new planets corresponding to radial
velocity signals, which are not sensitive to the choice of noise
  models and are identified in multiple data sets over various
  timespans. Eight candidate planets require additional observation to
  be confirmed. We also confirm three previously reported
  planets. Among the new planets, GJ 180 d and GJ 229A c are
  super-Earths located in the conservative habitable zones of their host stars. We investigate their dynamical stability using the
Monte Carlo approach and find both planetary orbits are robust to the
gravitational perturbations of the companion planets. Due to their
proximity to the Sun, the angular separation between the host stars
and the potentially habitable planets in these two systems is 25 and
59\,mas, respectively. They are thus good candidates for future direct
imaging by JWST and E-ELT. In addition we find GJ 433 c, a cold
super-Neptune belonging to an unexplored population of
Neptune-like planets. With a separation of 0.5\,as from its host star,
GJ 433 c is probably the first realistic candidate for direct imaging of cold Neptunes. A comprehensive survey of these planets is important for the studies of planet formation. 
\end{abstract}

\section{Introduction}\label{sec:intro}
The precision Doppler velocity revolution began in the early 1980s
\citep{campbell79, campbell88} but proceeded slowly through the
mid 1990s.  Prior to the 1980s Doppler velocity precision had been
stalled at 300 \ms for many decades, spanning the photographic, early
digital, and CCD eras. By achieving a long term precision of 13 \ms
the Campbell-Walker team improved Doppler precision by nearly two
orders of magnitude. They also demonstrated that some sun-like star were
intrinsicially stable enough to potentially pursue measurements at
higher precision.

In the era before exoplanets, Jupiter was the
benchmark. Jupiter gravitationally induces a 12 \ms velocity variation
on the Sun. A convincing 3-to-4 sigma detection of a Jupiter-analog
requires precision of 3 \ms. Over the past 30 years two techniques
have generated most of the improvement in Doppler velocity measurement
precision. The first exoplanet was found by the stabilized spectrometer method
\citep{mayor95}. The next 14 planets were found by the Iodine
absoprtion cell technique (\citealt{butler06}, Table 3). The Iodine
technique first achieved a precision of 3 \ms in 1995
\citep{butler96}, and the stabilized spectrometer first achieved a
precision of 1 \ms in 2004 \citep{rupprecht04,pepe11}. The data sets reported in this paper come from both techniques.

Due to their lower mass, M dwarfs are the primary class of stars
for which terrestrial mass planets can be found via the precision
Doppler technique. The first planet in the terrestrial mass
regime was found around the M dwarf GJ 876 \citep{rivera05}.
Over the past decade M dwarfs have been the principle targets
for potentially habitable planets \citep{vogt10,anglada16,astudillo16,feng17c} because their habitable zones are much closer to the star, and thus the potentially habitable planets have much shorter periods (and in turn produce larger semi-amplitudes) than those orbiting around G stars.

The Ultraviolet and Visual Echelle Spectrograph (UVES) M Dwarf Planet survey included 33 stable nearby
M dwarfs (\citealt{zechmeister09}; herafter ZKE2009).
\cite{butler19} (hereafter B19) have reanalyzed this data set, starting with
the raw images from the ESO archive \footnote{\url{archive.eso.org}}.
The updated UVES M dwarf data have previously contributed to the
discovery of the terrestrial mass planets around proxima Cen
\citep{anglada16} and Barnard's star \citep{ribas18}.

Velocity data sets from UVES, the High Accuracy Radial velocity Planet
Searcher (HARPS), the Carnegie Planet Finder Spectrograph (PFS)
mounted on Magellan, and the High Resolution Echelle Spectrometer
(HIRES) mounted on Keck have been combined to search for periodicities and confirm signals found
in the newly reanalyzed UVES data. Sixteen planet candidates
have been found orbiting nine nearby M dwarfs.  Of these three have
previously been announced (\citealt{tuomi14}; \citealt{nakajima95}),
five are newly announced low-mass planets, and eight remain candidates requiring
more observations. Three additional stars with previously announced
planets are also discussed.

This paper is the second of a series aiming at finding Earth analogs and the
methodology in this paper is similar to that in \cite{feng19a}
(hearafter Paper I). Section \ref{sec:data} will describe the stars
and the velocity data sets.  Section \ref{sec:results} will examine
the data sets for periodicities and embedded planetary signals.
Conclusions will be presented in Section \ref{sec:conclusion}.

\section{Radial Velocity Observations} \label{sec:data}

The physical and observational properties for the stars in this
study are listed in Table \ref{tab:star}.  These stars are drawn
from the recently reanalyzed data from the UVES M Dwarf Planet
Search (ZKE2009; B2019). The first two columns of the table list
common catalog designations for the stars.  The spectral type, stellar
mass,  and V magnitudes are shown in the third, fourth and fifth
columns respectively. These are from table 2 of ZKE2009, and also table 1 of
B19.  

Columns 6 through 11 list the number of observations taken with each
spectrometer. HARPS and PFS have each had one major upgrade.  In May
2015 the fiber that feeds HARPS was replaced.  In January 2018 the
old PFS CCD (4Kx4K, 15 micron pixels) was replaced with a next
generation CCD (10Kx10K, 9 micron pixels).  Simultaneously
the PFS default slit width for iodine observations was reduced from
0.5$''$ to 0.3$''$, increasing the resolving power of the instrument from $\sim$80K to $\sim$130K. 

\begin{table}
\caption{Stellar parameters and radial velocity (RV) data sets
  for stars. The values of stellar types and masses are from
  ZKE2009. The number of RV points are shown for the UVES (U), KECK
  (K), HARPSpre (H1), HARPSpost (H2),PFSpre (P1) and PFSpost (P2) data set.}
\label{tab:star}
\centering
\begin{tabular}{*{11}{c}}
\hline\hline
Star name&Other Name&Spectral type& V (mag)&Stellar mass ($M_\odot$)&U&K&H1&H2&P1&P2\\\hline
  GJ 27.1 &HIP 3143&M0.5&11.42&0.53&62&8&50&0&0&0\\
  GJ 160.2&HIP 19165&M0V&9.69&0.69&101&45&44&0&15&3\\
  GJ 173 &HIP 21556&M1.5&10.35&0.48&12&34&16&0&0&0\\
  GJ 180 &HIP 22762&M2V&12.50&0.43&57&64&78&9&49&0\\
  GJ 229A &HD 42581&M1/M2V&8.14&0.58&74&47&124&76&0&0\\
  GJ 422 &HIP 55042&M3.5&11.66&0.35&24&0&45&7&0&0\\
  GJ 433 &HIP 56528&M1.5&9.79&0.48&167&33&86&0&39&4\\
  GJ 620 &HIP 80268&M0&10.25&0.61&5&0&23&0&0&0\\
  GJ 682 &HIP 86214&M3.5V&10.96&0.27&49&0&20&0&0&0\\
  GJ 739 &HIP 93206&M2&11.14&0.45&48&0&19&0&0&0\\
  GJ 911 &HIP 117886&M0V&10.88&0.63&26&24&3&0&0&0\\
  GJ 3082 &HIP 5812&M0&11.10&0.47&10&0&42&0&0&0\\\hline
\end{tabular}
\end{table}

All of the stars in Table \ref{tab:star} were observed with UVES
on the VLT-UT2 telescope between March 2000 and March 2007
(ZKE2009). We introduce UVES as well as other RV data used in this work as follows.

\begin{itemize}
\item {\bf UVES}
UVES is a dual arm cross dispersed echelle spectrometer
\citep{dekker00}.  Though perhaps not fully appreciated at the time,
UVES was the first ``modern'' precision velocity instrument. The
most important criterion for a precision velocity spectrometer
is resolution. With a resolution of 130\,K, UVES operates at twice
the resolution of earlier echelles. The calibration for precision
velocity measurements is provided by an Iodine absorption
cell \citep{marcy92}.

The UVES M dwarf group used the ``AUSTRAL'' Iodine code to model
the observed spectra and produce Doppler velocity measurements
\citep{endl00}. The AUSTRAL code is based on the the
modeling process outlined in \cite{butler96}.  The resulting median
velocity RMS of the 33 stable stars is 5.5 \ms. 

The UVES data has been re-reduced using a custom raw reduction
package, and an upgraded version of code from \cite{butler96}.
The median velocity RMS of the stable stars is reduced to 3.6 \ms
(B19).  The velocities reported here are the ``unbinned''
velocities published in B19.
\item {\bf HARPS}
HARPS has been the premier precision velocity instrument since 
its inception \citep{rupprecht04}, routinely
approaching or exceeding a precision of 1 \ms \citep{pepe11}.
We have obtained all the publicly reduced HARPS spectra from
the ESO archive and generated velocities with the HARPS-TERRA
package \citep{anglada12}.

\item {\bf PFS}
  PFS \citep{crane10} is a purpose built Iodine precision velocity
echelle that is used on the 6.5-m Magellan II (Clay) telescope.
Like HARPS it is designed to maximize thermal and mechanical
stability.  With the exception of the focus, PFS has no moving parts.
It is mounted on an optical bench in a thermal insulating enclosure.
The interior of the enclosure is heated to 27 \deg C, and the temperature
is maintained to $\pm$ 0.01 \deg C.  The focus of PFS does not change
with time.  We have reduced the PFS data with our custom raw and
velocity reduction packages.

\item {\bf KECK}
The Keck HIRES program is the longest continously running precision
velocity survey, having commenced in 1996. HIRES \citep{vogt94} is
permanenly mounted on the nasmyth platform on the Keck I 10-m
telescope.  An Iodine cell is used for the wavelength calibration.
Most of the first 200 extrasolar planets were found with this system
\citep{butler06}. With a 0.86$''$ slit, the resolution of HIRES is 60\,K. As a result the long term precision of HIRES is 2-to-3
\ms. The data for this program is from the analysis of
  \cite{butler17}. We have reduced all the data, starting with the
raw images, with our custom raw and velocity reduction packages.
\end{itemize}

\begin{figure*}
  \centering
  \includegraphics[scale=0.5]{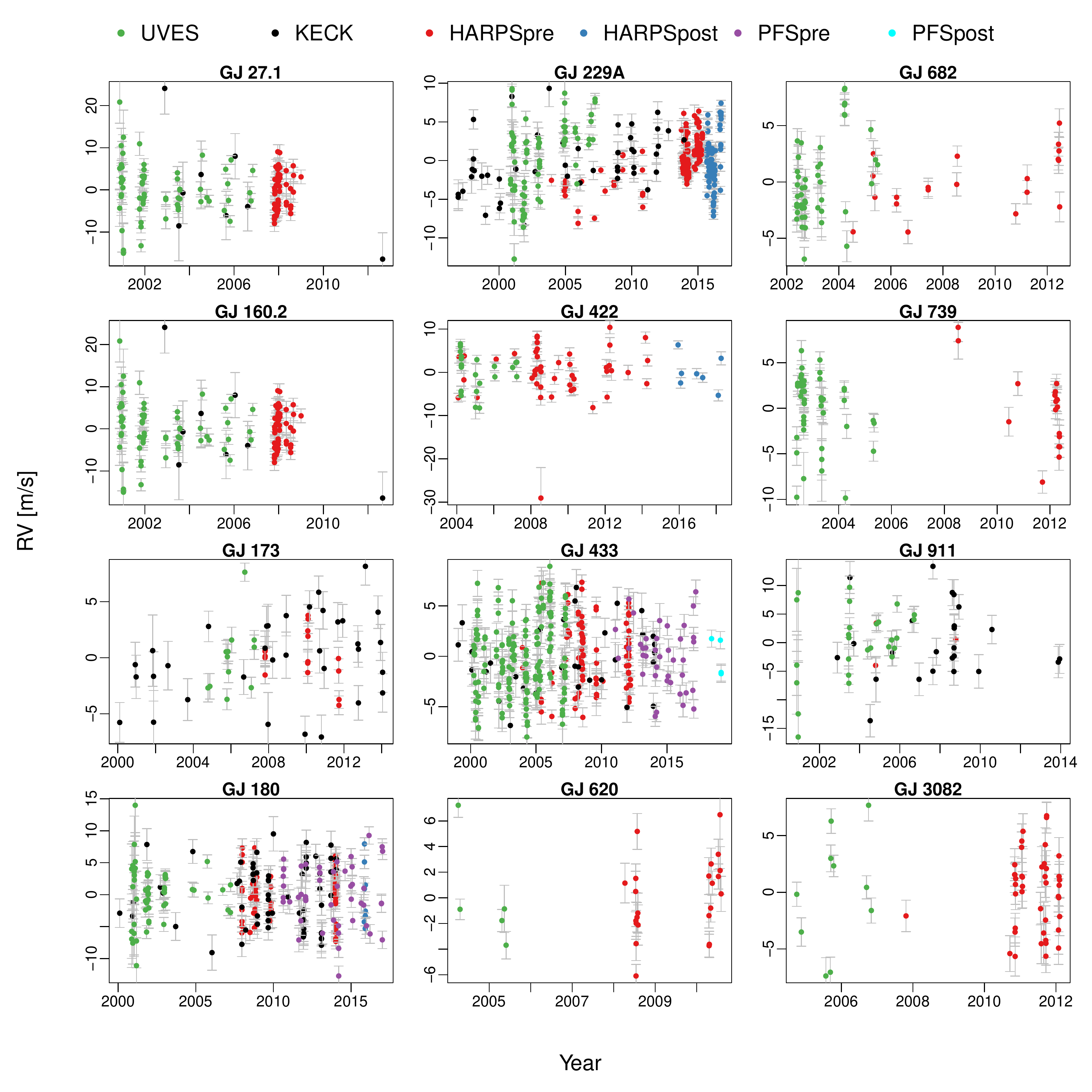}
  \caption{Doppler velocity data sets of UVES, KECK, HARPS and PFS
    for the 12 stars reported in this paper. The RV sets are
      shifted to zero mean for optimal visualization.}
  \label{fig:data}
\end{figure*}

Nine of the stars listed in Table \ref{tab:star} are found to host
new planets. The remaining three stars have previously announced
planets (\citealt{tuomi14}; hereafter T14). We are able to confirm
some of the planets from T14, but not all of them. We discuss this discrepancy in section \ref{sec:T14}. Figure \ref{fig:data} shows the full
Doppler velocity data sets for the 12 stars. There is significant
temporal overlap between the data sets for many of the stars. 


\section{Method}\label{sec:method}
The method used in RV data analysis in this work is similar to
  that used in Paper I. We briefly introduce it in this section. 

\subsection{RV model and model selection}\label{sec:model}
The RV model is composed of the signal and noise components. The
  signal component for $N_p$ planets for the $k^{\rm th}$ data set is
  \begin{equation}
    \hat{v}^k_s(t_j)=\sum_{i=1}^{N_p}
    K_i\left[\sin{(\omega_i+\nu_i(t_j))}+e_i\cos{\omega_i}\right]+\gamma_k+\dot{\gamma}_kt_j~,
  \end{equation}
  where $K_i$ is the semi-amplitude of stellar RV
  variation caused by the perturbation of the $i^{\rm th}$ planet,
  $\nu_i(t_j)$ is the true anomaly derived from the orbital period
 $P_i$, eccentricity $e_i$ and the reference mean anomaly $M_0$ by
 solving Kepler's equation, $\gamma_i$ and $\dot{\gamma}_i$ are
 respectively the intercept and slope of a linear trend used to model
 instrumental bias and secular acceleration. For long period signals
 with period comparable with the RV data timespan, we replace the linear
 trend by an offset to avoid degeneracy between long period signals
 and linear trend. 

The time-correlated (or red) noise in the $k^{th}$ RV set is modeled by
the $q^{\rm th}$-order moving average model (MA(q)), 
  \begin{equation}
    \hat{v}^k_{\rm n}(t_j)=\sum_{i=1}^{q}w_i^k{\rm exp}(-\frac{|t_j-t_{j-i}|}{\tau^k})[v^k_{j-i}-\hat{v}_s^k(t_{j-i})]~,
    \end{equation}
where $w_i^k$ is the amplitude of the $i^{th}$ MA component for the
$k^{\rm th}$ RV set, and $\tau^k$ is the time scale of the MA model
for the $k^{\rm th}$ RV set, and $v^k_{j-i}$ is the measured RV at epoch
$t_{j-i}$ in the $k^{\rm th}$ set. The full RV model for the $k^{th}$ RV set is
\begin{equation}
  \hat{v}^k(t_j)=\hat{v}^k_s(t_j)+ \hat{v}^k_{\rm n}(t_j)~.
\end{equation}
The logarithmic likelihood for the combination of $N_s$ RV sets is
\begin{equation}
\ln{\mathcal{L}}=-\frac{1}{2}\sum_{k=1}^{N_s}\sum_{j=1}^{N_k}\ln{[2\pi
  (\sigma_k^2(t_j)+s_k^2)]}-\frac{1}{2}\sum_{k=1}^{N_s}\sum_{j=1}^{N_k}\frac{[v^k_j-\hat{v}^k(t_j)]^2}{\sigma_k^2(t_j)+s_k^2}~,
\end{equation}
where $s_k$ is white noise jitter for the $k^{\rm th}$ RV set and $\sigma_k(t_j)$ is the measured
RV uncertainty at epoch $t_j$ in the $k^{\rm th}$ RV set, $N_k$ is the
number of RV points in set $k$.

The difference in the Bayesian
information criterion (BIC) of two models is 
\begin{equation}
  \Delta{\rm BIC}_{12}=\ln{\mathcal{L}^{\rm max}_2}-\ln{\mathcal{L}^{\rm
        max}_1}+(n_1-n_2)\ln N_{\rm rv}~,
  \end{equation}
  where $\mathcal{L}^{\rm max}_1$ and $\mathcal{L}^{\rm max}_2$ are
  respectively the maximum likelihoods of model 1 and 2, $N_{\rm
  rv}$ is the number of all RV points for a target, $n_1$ and $n_2$
are respectively the efficient numbers of free parameters in model 1
and 2. 
Following \cite{kass95}, we convert $\Delta{\rm BIC}_{\rm 12}$ to
Bayes factor (BF) by assuming a single Gaussian posterior
distribution, $\ln{\rm BF}_{21}=\frac{1}{2}\Delta{\rm
  BIC}_{12}$. This assumption is not inappropriate as long as the
posterior is dominated by a single signal. Moreover, the threshold $\ln{\rm BF}_{21}>5$ or $\Delta{\rm BIC}_{12}>10$ are appropriate for model selection according to
\cite{kass95}. This is also confirmed by a comparison of various information
criteria and BF computation methods based on analyses of
synthetic and real RV sets in \cite{feng16}. The order $q$ of the MA
model is determined through Bayesian model comparison by selecting the
model with the highest order which passes the $\ln{\rm BF}_{21}>5$
criterion. Hereafter we use $\ln{\rm BF}$ as an abbreviation of
$\ln{\rm BF}_{21}$. 

\subsection{Posterior sampling and noise model selection}\label{sec:posterior}
We use adaptive Markov Chain Monte Carlo (MCMC) developed by
  \cite{haario06} to sample the posterior distribution of model
  parameters. We adopt a semi-Gaussian prior ($P(e)=\mathcal{N}(0,0.2)~
  \forall e \ge 0$) for eccentricity in order to capture the broad
  feature of eccentricity distribution found in Kepler planet samples
  \citep{kane12,VanEylen18}, a logarithmic uniform prior for orbital
  period and MA time scale, and uniform priors for other
  parameters. Since we have adopted an informative prior for
  eccentricity, we test the sensitivity of our results to
  eccentricitiy priors for strong planetary candidates in section \ref{sec:prior} and do not find
  significant dependence of parameter values on priors although minor sensitivity might be found for weak planetary candidates. 

For a given model, we sample the posterior through multiple tempered
(hot) MCMC chains to identify the global maximum of posterior. We then
use non-tempered (cold) chains to sample the global maximum found by
hot chains. From the posterior sample, we infer the parameter at the
maximum a posteriori (MAP) and use quantiles to estimate parameter
uncertainties. This is explained in detail in Paper I. 

To select the optimal noise model, we calculate the maximum likelihood
for a MA model using the Levenberg-Marquardt (LM) optimization
algorithm \citep{levenberg44,marquardt63}. We calculate ln(BF) for
MA(q+1) and MA(q). If ln(BF)$<5$, we select MA(q). If ln(BF)$\ge 5$,
we select MA(q+1) and keep increasing the order of MA model until
the model with the highest order passing the ln(BF)$\ge 5$ criterion
is found. Readers are referred to \cite{feng17b} for details. 

\subsection{Signal selection criteria}\label{sec:criteria}
Following Paper I, we select signals which are statistically
significant, independent of noise models, not correlated with
stellar activity and consistent in time. Here we reiterate the main points of these criteria which are described in Paper I.

The noise models we have used to calculate Bayes factor
periodograms (BFPs; \citealt{feng17b}) are the white noise model (a
constant jitter is used to fit excess noise), the first order moving
average model (MA(1); \citealt{tuomi12}), and the first
auto-regressive model (AR(1); \citealt{tuomi13}). The evidence for signal is considered to be strong if the logarithmic Bayes factor is larger than 3 (i.e. $\ln{\rm BF}>3$) or equivalently its Bayesian
information criterion (BIC) is larger than 6 \citep{kass95}. A signal
is considered to be very strong or significant if $\ln{\rm
  BF}>5$. However, the exact number of free parameters $k$ is not
known. A Keplerian model for a circular orbit has three free
parameters while an eccentric orbit needs five parameters to
model. Hence we define $k=3$ and $k=5$ as the boundaries for the real
BIC value, corresponding to $\ln{\rm BF}_3$ and $\ln{\rm BF}_5$,
respectively. Signals with $\ln{\rm BF}_3>5$ are selected as planet
candidates. Since the sinusoidal function is used in the calculation of
  BFPs, we use $\ln{\rm BF}_3=5$ as a threshold to visualize the
  significance of a signal. The real significance of a signal is determined through posterior sampling combined with the BF threshold.

To exclude signals due to stellar activity, we calculate BFPs
for activity indices and window functions for each data set and find whether there is an overlap between RV signals and activity
signals. To assess the consistency of signals over time, we show
the moving periodogram for those signals whose phase is well covered
by the RV data. Specifically, the BFP is calculated for the RVs
  measured within a time window. The time window moves with a certain
  time step until the whole timespan is covered. The BFPs for all time windows
  form a two dimensional map of periodogram powers. Considering that the number
  of RVs in a time window changes when the window moves, the BFP for a
  given step is normalized so that the power varies from 0 to 1. For
  signals with orbital period comparable with the data timespan, one
  should not rely on the moving periodogram as the single diagnostic
  tool. Since the RVs are typically not measured in a uniform way, the
  consistency of a true signal may depend on the sampling cadence
  even if the power is normalized (Paper I). However, it is easy to
  identify false positives if inconsistency is found at high cadence
  epochs with a timescale comparable to or longer than the signal period.

The moving periodogram is also known as
  time-frequency analysis in analyses of regularly spaced time
  series \citep{cohen95}. Compared with time-frequency
  analysis, the moving periodogram accounts for floating linear trend, red
  noise, jitter and irregularity in RV data \citep{feng17b}. Thus we
  follow \cite{feng17b} by using ``moving periodogram'' to distinguish between these two types of analyses. Similar techniques have been developed by
  \cite{mortier15,mortier17} to test the sensitivity of signals to
  sample size.

\section{Results}\label{sec:results}

\subsection{Planetary signals}
We select those signals which have $\ln{\rm BF}_3>5$, do not display
significant activity, and are independent of the chosen noise model. We show the orbital solutions for these signals in Table
\ref{tab:planet}. Notes are attached for signals to show whether they
are temperate and to show concerns about their quality. Planet
candidates with notes of ``OV'', ``L3'', ``NC'' are to be further
confirmed while candidates with ``HZ'' or without any notes are likely
to be real planets. Based on these considerations, we find eight
planet candidates including two reported by T14, which need to be
confirmed. They are GJ 173 b, GJ 229A b, GJ 433 c, GJ 620 b, GJ
620 c, GJ 739 b, GJ 739 c, and GJ 911 b. 

Our analyses support a Keplerian origin for GJ 180 b and GJ 433 b
detected by T14 as well as GJ 229B discovered by \cite{nakajima95}. We
identify five new planets corresponding to RV signals which are
statistically significant, unique and consistent over time, and robust
to the choice of noise models. They are GJ 180 d, GJ 229A c,
GJ 422 b, GJ 433 d, and GJ 3082 b. Two of the five new
planets have masses in line with super-Earth type planets located in
the habitable zones of their hosts.

\begin{figure*}
  \hspace{-0.3in}\includegraphics[scale=0.5]{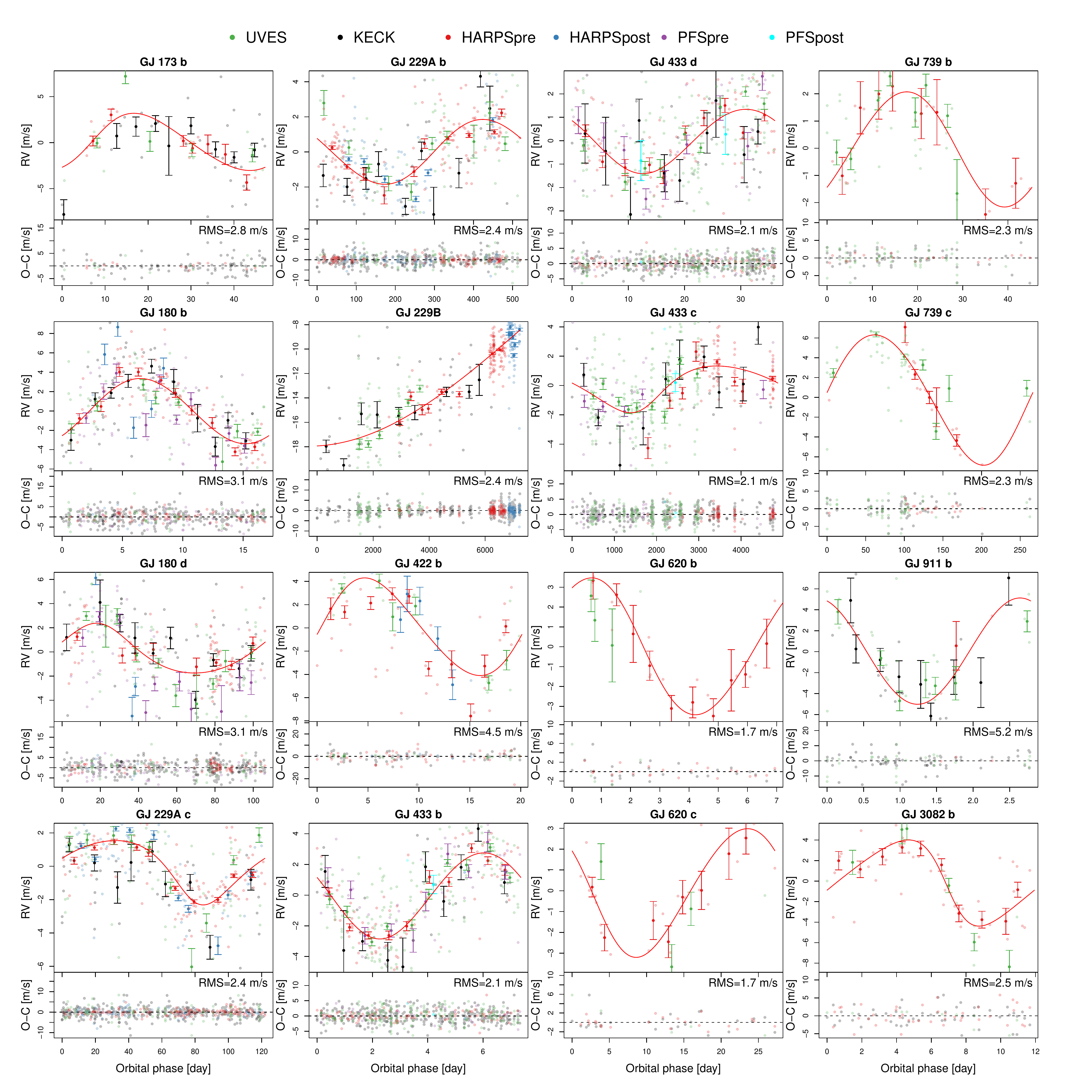}
  \caption{Phase curves and corresponding residuals for all planet candidates are shown. The instruments are encoded by different colors and
    are shown on top of all panels. The error bars show the
    error-weighted average RVs in 10 evenly separated time bins. The
    best orbital solution is determined by the MAP values of orbital parameters. The Root Mean Square (RMS) of the residual RVs
after subtracting all signals is shown in each panel. }
  \label{fig:phase}
\end{figure*}

We show the phase curve and residuals for all planet candidates in Fig. \ref{fig:phase}. Since the residuals probably contain insignificant planetary and activity signals, the
  traditional test of goodness of fit may not be suitable for this
  type of residuals. Nevertheless we report the results of various
  tests in Table \ref{tab:tests} and discuss their implications for the case of irregularly and sparsely sampled RV time series.

\begin{table}
\caption{Parameters for planet candidates. The minimum mass, semi-major axis,
  period, RV semi-amplitude, eccentricity, and mean anomaly at the
  reference epoch are denoted by $M_p\sin{I}$, $a$, $P$, $K$, $e$,
  $\omega$, and $M_0$, respectively. The mean and standard deviation
  of each parameter are estimated from the posterior samples drawn by
  MCMC. For each parameter, the value at the MAP and the uncertainty
  interval defined by the 1\% and 99\% quantiles of the posterior
  distribution are shown below the values of mean and standard
  deviation. The note for a planet candidate shows that it is in the habitable zone (HZ) defined by \cite{kopparapu14}, or it is
  reported in T14 or in \cite{nakajima95} (N95), or it partly overlaps
  with activity signals (``OV''), or $\ln{\rm BF}_5$ is less
  than 3 (``L3''), or it is not found in individual data sets (NI). If a signal is not consistently significant over time due to a lack of enough data, we add note ``NC'' to show our concern. For the planet candidates reported by T14, we keep their original names
 and assign new names to new planet candidates identified in this
 work. We use bold-faced planet name to indicate a reliable detection
 of planet, use italic font to indicate planet candidates needing to be confirmed by more observations and use normal font to indicate
 confirmation of previously reported signals.}
\label{tab:planet}
\hspace{-0.5in}
\begin{tabular}{*9{l}}
\hline\hline
Planet&$M_p\sin{I}$ ($M_\Earth$)& $a$ (au) & $P$ (day) &$K$
                                                         (m/s)&$e$&$\omega$
                                                                    (deg) &  $M_0$ (deg) &Note\\\hline
{\it GJ 173
  b}&$9.5\pm2.1$&$0.200\pm0.007$&$47.304\pm0.254$&$2.78\pm0.59$&$0.10\pm0.06$&$231\pm94$&$222\pm106$&NI\\
&$10.70_{-5.92}^{+3.96}$&$0.200_{-0.017}^{+0.015}$&$47.260_{-0.543}^{+0.639}$&$3.11_{-1.94}^{+0.93}$&$0.12_{-0.11}^{+0.16}$&$283_{-279}^{+72}$&$297_{-294}^{+60}$&\\\hline
GJ 180 b&$6.49\pm0.68$&$0.092\pm0.003$&$17.133\pm0.003$&$3.25\pm0.26$&$0.07\pm0.04$&$276\pm26$&$314\pm25$&T14\\
&$6.75_{-1.78}^{+1.37}$&$0.092_{-0.008}^{+0.007}$&$17.132_{-0.006}^{+0.008}$&$3.37_{-0.61}^{+0.47}$&$0.02_{-0.01}^{+0.14}$&$255_{-17}^{+78}$&$327_{-75}^{+28}$&\\\hline
{\bf GJ 180 d}&$7.56\pm1.07$&$0.309\pm0.010$&$106.300\pm0.129$&$2.08\pm0.26$&$0.14\pm0.04$&$155\pm127$&$235\pm85$&HZ\\
&$7.49_{-2.33}^{+2.66}$&$0.310_{-0.026}^{+0.022}$&$106.341_{-0.340}^{+0.261}$&$2.06_{-0.43}^{+0.58}$&$0.16_{-0.07}^{+0.06}$&$10_{-8}^{+345}$&$292_{-283}^{+53}$&\\\hline
{\bf GJ 229A c}&$7.268\pm1.256$&$0.339\pm0.011$&$121.995\pm0.161$&$1.93\pm0.30$&$0.19\pm0.08$&$121\pm33$&$154\pm36$&HZ\\
&$7.93_{-3.44}^{+2.39}$&$0.339_{-0.029}^{+0.024}$&$122.005_{-0.382}^{+0.364}$&$2.15_{-0.89}^{+0.50}$&$0.29_{-0.26}^{+0.06}$&$101_{-64}^{+104}$&$172_{-95}^{+88}$\\\hline
{\it GJ 229A b}&$8.478\pm2.033$&$0.898\pm0.031$&$526.115\pm4.300$&$1.37\pm0.31$&$0.10\pm0.06$&$199\pm71$&$212\pm82$&OV,T14\\
&$10.02_{-6.10}^{+3.35}$&$0.896_{-0.073}^{+0.069}$&$523.242_{-7.009}^{+13.039}$&$1.63_{-1.02}^{+0.38}$&$0.17_{-0.17}^{+0.07}$&$283_{-275}^{+65}$&$160_{-159}^{+191}$&\\\hline
GJ 229B&$426.389\pm57.505$&$17.585\pm2.477$&$45925.334\pm9473.054$&$15.48\pm1.47$&$0.07\pm0.05$&$199\pm93$&$206\pm114$&N95\\
&$514.79_{-211.31}^{+57.62}$&$19.433_{-6.893}^{+4.710}$&$52890.273_{-25055.401}^{+19639.584}$&$17.66_{-5.20}^{+1.28}$&$0.03_{-0.03}^{+0.21}$&$180_{-179}^{+176}$&$17_{-16}^{+340}$&\\\hline
{\bf GJ 422 b}&$11.07\pm1.12$&$0.111\pm0.004$&$20.129\pm0.005$&$4.47\pm0.34$&$0.11\pm0.04$&$265\pm18$&$224\pm160$&\\
&$10.44_{-1.87}^{+3.33}$&$0.111_{-0.009}^{+0.008}$&$20.129_{-0.012}^{+0.012}$&$4.20_{-0.35}^{+1.00}$&$0.10_{-0.10}^{+0.12}$&$283_{-54}^{+12}$&$342_{-342}^{+18}$&\\\hline
GJ 433 b&$6.043\pm0.597$&$0.062\pm0.002$&$7.3705\pm0.0005$&$2.86\pm0.21$&$0.04\pm0.03$&$154\pm106$&$170\pm114$&T14\\
&$5.94_{-1.23}^{+1.55}$&$0.062_{-0.005}^{+0.004}$&$7.3708_{-0.0014}^{+0.0009}$&$2.81_{-0.44}^{+0.51}$&$0.02_{-0.02}^{+0.10}$&$140_{-139}^{+216}$&$287_{-284}^{+69}$&\\\hline
{\bf GJ 433 d}&$5.223\pm0.921$&$0.178\pm0.006$&$36.059\pm0.016$&$1.46\pm0.24$&$0.07\pm0.05$&$159\pm82$&$219\pm102$&\\
&$4.94_{-1.79}^{+2.52}$&$0.178_{-0.015}^{+0.013}$&$36.052_{-0.031}^{+0.045}$&$1.37_{-0.51}^{+0.66}$&$0.03_{-0.02}^{+0.16}$&$154_{-143}^{+194}$&$258_{-256}^{+97}$&\\\hline
{\it GJ 433 c}&$32.422\pm6.329$&$4.819\pm0.417$&$5094.105\pm608.617$&$1.75\pm0.31$&$0.12\pm0.07$&$242\pm55$&$213\pm51$&L3,T14\\
&$28.78_{-10.46}^{+19.15}$&$4.692_{-0.768}^{+1.169}$&$4873.923_{-1034.762}^{+1796.128}$&$1.60_{-0.56}^{+0.88}$&$0.21_{-0.21}^{+0.08}$&$218_{-180}^{+126}$&$233_{-145}^{+105}$&\\\hline
{\it GJ 620 b}&$7.26\pm1.16$&$0.063\pm0.002$&$7.655\pm0.004$&$3.41\pm0.49$&$0.08\pm0.05$&$182\pm110$&$180\pm107$&NC\\
&$7.38_{-2.70}^{+2.68}$&$0.063_{-0.005}^{+0.005}$&$7.652_{-0.006}^{+0.012}$&$3.45_{-1.31}^{+1.03}$&$0.06_{-0.05}^{+0.16}$&$84_{-80}^{+272}$&$251_{-247}^{+106}$\\\hline
{\it GJ 620 c}&$6.97\pm2.34$&$0.147\pm0.005$&$27.040\pm0.429$&$2.15\pm0.70$&$0.09\pm0.06$&$154\pm110$&$204\pm108$&NC,L3\\
&$10.08_{-8.46}^{+2.48}$&$0.148_{-0.013}^{+0.011}$&$27.219_{-1.164}^{+0.828}$&$3.09_{-2.68}^{+0.53}$&$0.09_{-0.08}^{+0.15}$&$115_{-112}^{+242}$&$303_{-298}^{+55}$&\\\hline
{\it GJ 739 b}&$9.75\pm1.86$&$0.211\pm0.007$&$45.357\pm0.043$&$2.45\pm0.44$&$0.08\pm0.06$&$169\pm104$&$178\pm99$&NC\\
&$8.49_{-2.90}^{+5.79}$&$0.211_{-0.018}^{+0.015}$&$45.323_{-0.067}^{+0.134}$&$2.12_{-0.71}^{+1.46}$&$0.04_{-0.04}^{+0.21}$&$120_{-120}^{+233}$&$104_{-101}^{+252}$\\\hline
{\it GJ 739 c}&$42.63\pm6.23$&$0.687\pm0.023$&$266.985\pm0.745$&$5.91\pm0.77$&$0.07\pm0.04$&$197\pm88$&$150\pm113$&NC\\
      &$47.52_{-18.82}^{+10.13}$&$0.687_{-0.057}^{+0.051}$&$266.489_{-1.232}^{+2.232}$&$6.58_{-2.25}^{+1.59}$&$0.06_{-0.06}^{+0.13}$&$228_{-225}^{+121}$&$43_{-42}^{+315}$&\\\hline
{\it GJ 911 b}&$6.9\pm1.3$&$0.033\pm0.001$&$2.7889\pm0.0004$&$4.31\pm0.73$&$0.08\pm0.06$&$181\pm108$&$184\pm103$&NC\\
&$8.21_{-4.12}^{+1.78}$&$0.033_{-0.003}^{+0.002}$&$2.7888_{-0.0008}^{+0.0009}$&$5.08_{-2.49}^{+0.94}$&$0.01_{-0.01}^{+0.24}$&$26_{-24}^{+330}$&$353_{-349}^{+2}$&\\\hline
{\bf GJ 3082 b}&$8.2\pm1.7$&$0.079\pm0.003$&$11.949\pm0.022$&$3.94\pm0.74$&$0.22\pm0.11$&$121\pm54$&$150\pm56$&\\
&$8.77_{-4.25}^{+3.50}$&$0.079_{-0.007}^{+0.006}$&$11.942_{-0.043}^{+0.060}$&$4.20_{-2.00}^{+1.36}$&$0.26_{-0.25}^{+0.23}$&$100_{-82}^{+240}$&$148_{-124}^{+186}$&\\
\hline
\end{tabular}
\end{table}

\begin{figure*}
  \centering
  \includegraphics[scale=0.4]{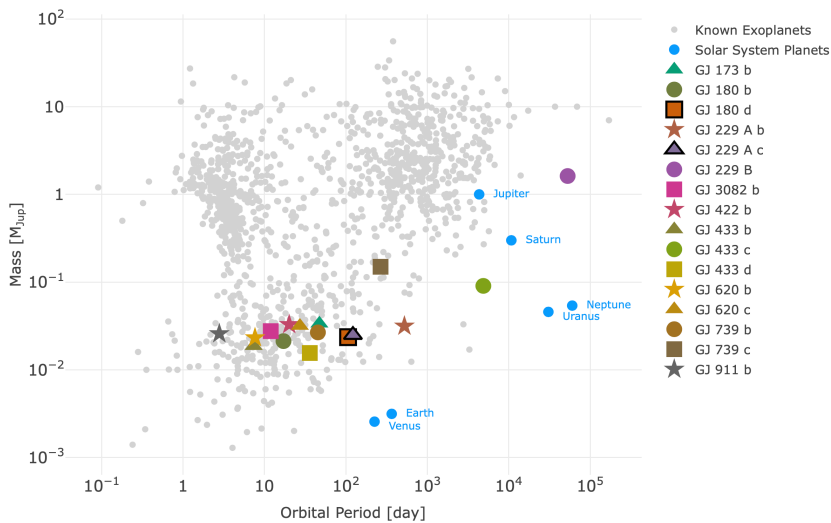}
  \caption{Distribution of planet mass and period for archived exoplanets, Solar System planets and the ones found in this work. The shapes and colors of markers are combined to show different planets. The two habitable-zone planets, GJ 180 d and GJ 229 A c, are denoted by black borders.}
  \label{fig:mp}
\end{figure*}

In Fig. \ref{fig:mp}, we visualize the distribution of planet mass and
period for the planet candidates detected in this work and the planets
collected by NASA Exoplanet
Archive\footnote{\url{https://exoplanetarchive.ipac.caltech.edu/index.html}} \citep{akeson13}. While
most planets are located in the crowded region in the parameter space,
GJ 433 c is a Neptune on a year-long orbit. The detection of
wide-orbit Neptunes around M dwarfs has thus far been rare. Moreover,
GJ 433 c stands out as a unqiue detection of a wide-orbit cold
super-Neptune. Previous detections of cold Neptunes have come from
the microlensing technique (e.g., \citealt{sumi10}). Our confirmation
of the previously suspected super-Neptune around GJ 433 (T14)
demonstrates the ability of increasingly precise Doppler velocity
measurements and longer Doppler baselines to probe this rarely explored population.

\begin{figure*}
  \centering
      \includegraphics[scale=0.6,angle=-90]{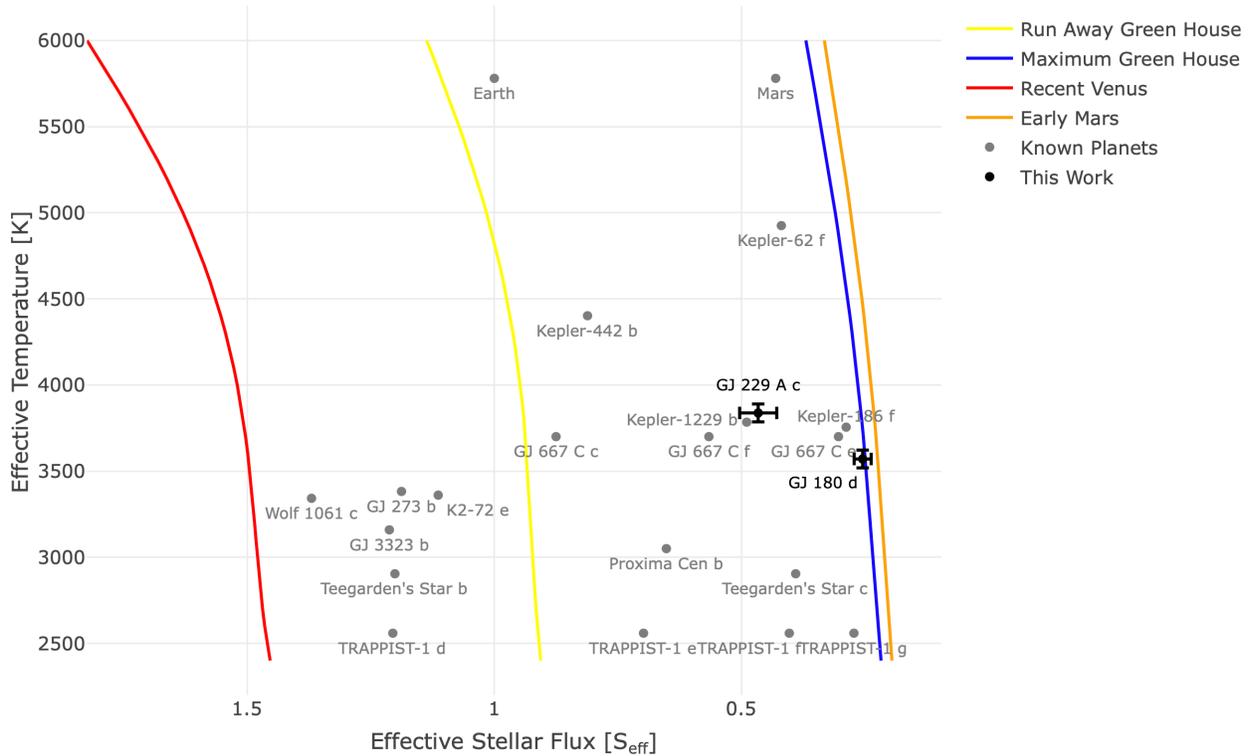}
  \caption{Distribution of incoming stellar flux and stellar efficient temperature for known planets and planets detected in this work. The conservative HZ is defined by the run away green house and maximum green house limit \cite{kopparapu14}. The optimistic HZ is defined by the recent Venus and early Mars limit. The effective temperature and luminosity of GJ 229A c and GJ 180 d are from \cite{schweitzer19}. The parameters for exoplanets are from the NASA Exoplanet Archive. The error bars for GJ 229A c and GJ 180 d are uncertainties of effective temperature and effective stellar flux due to measurement errors in stellar mass and luminosity. }
  \label{fig:hz}
\end{figure*}

We compare the two temperate planets, GJ 180 d and GJ 229A c, with
the conservative sample of potentially habitable exoplanets collected
by PHL\footnote{\url{http://phl.upr.edu/projects/habitable-exoplanets-catalog}}
in Fig. \ref{fig:hz}. GJ 229A c is located within the
conservative HZ defined by \cite{kopparapu14} while GJ 180 d is
located near the HZ outer edge corresponding to the maximum green
house limit. GJ 180 d is in a wider orbit and thus receives less
stellar radiation than GJ 229A c. However, GJ 180 d and GJ 229A c
have minimum masses of 7.6 and 7.2\,$M_\oplus$.  Thus they might not be
rocky as the composition of super-Earths are not well
understood.

We discuss the results for each planet candidate in the following subsection. For each planet candidate, we show the BFPs for
RV data sets and activity indices. We label each panel with ``Pn'' where
n is a number to identify the panel. We also calculate the moving
periodogram to show the time consistency of signals. The components
in BFP figures are described in detail in the caption of
Fig. \ref{fig:BFP_GJ173}. The components in the moving periodogram figures
are introduced in Fig. \ref{fig:MP_GJ173b}. 

\subsection{Individual candidates}
\begin{itemize}
\item \textbf{GJ 173 (HIP 21556)} GJ 173 b has a minium mass of
  10.7\,$M_\oplus$ and an orbital period of 47.3 days. It is
  identified in the combined UVES, KECK and HARPS data (see Fig. \ref{fig:BFP_GJ173}). The activity indicators do not
  have signficant power at this period. In Fig. \ref{fig:MP_GJ173b},
  the moving periodogram shows consistent significance at the period
  of 47.3 days. However, this signal cannot be identified in
    individual data sets. Hence, we add a note in Table
    \ref{tab:planet} to show our concern.

\item \textbf{GJ 180 (HIP 22762)} GJ 180 b and d have minimum masses
  of 6.75 and 7.49\,$M_\oplus$, respectively. The corresponding RV
  signals with periods of 17 and 106 days are identified in the combined UVES, KECK, HARPS and PFS data. The 17-day signal reported by T14 is
  identified while the 24-day in T14 is not found to be
  significant, as shown in Fig. \ref{fig:BFP_GJ180}. There is no
  significant power at periods of 17 and 106 days in the BFPs for
  the activity indicators. In Fig. \ref{fig:MP_GJ180b} and
  \ref{fig:MP_GJ180c}, we see unique and consistent significance over
  time for these two signals. Therefore our analyses support a
  Keplerian origin of these two signals. The outer planet is a
  super-Earth located in the habitable zone. As the host star
  is only 12.4\,pc from the Sun, the separation between the primary and the potentially habitable planet is about 25\,mas.  

\item \textbf{GJ 229A (HD 42581)} There are three signals with periods
  of 122, 520 and 49,000 days found in the combined KECK, HARPSpre,
  HARPSpost and UVES data. They correspond to
  minimum masses of 7.93, 10.0, and 515\,$M_\oplus$. GJ 229A c is probably a super-Earth with
  an orbital period of 122 days and is located in the temperate zone
  corresponding to a period interval of $[114,315]$\,days
  \citep{kopparapu14}. As shown in Fig. \ref{fig:BFP_GJ229}, it does
  not overlap with any activity signals. It is strong in the
    HARPSpre, HARPSpost and UVES sets (e.g., P35, P39, P43 and P44). It is evident from the moving periodogram shown in Fig. \ref{fig:MP_GJ229b} that this signal is unique and is consistent over time although it is most significant in recent epochs due to high-cadence sampling. Although this signal is close to one third of one year, it is unlikely an annual alias because no strong powers around one year and half year periods are found in the BFPs nor in the moving periodograms. 

  The phase curve shown in Fig. \ref{fig:phase} demonstrates a good fit to the UVES and HARPS data. This strongly supports a Keplerian origin. This temperate super-Earth system is nearby (5.75 pc) and the separation between the primary and GJ 229 c is about 59\,mas, making it suitable for future direct imaging by facilities such as NIRSS on JWST \citep{greenhouse16} and EPICS on E-ELT \citep{marchiori08}. 

GJ 229A b has an orbital period of 523-day that is close to the period
of 470-day reported by T14 who used the old HARPS and UVES data
sets. The 471-day signal in T14 is probably related to the new 520
days signal found in our analysis. However, there are strong
powers around this period in the BFPs for activity indices shown in panel P57, P58, P59, P60,
P63, and P64 of Fig. \ref{fig:BFP_GJ229} although the peaks in the
activity index BFPs are quite broad. If we consider the activity
signal in P57 as typical, the long period activity signal probably
has a period between 200 and 2000 days. Thus any long period RV signal would
be identified as an activity signal if overlaps between RV and
activity signals are considered as the only diagnostic metric. Considering
the limitation of this criterion and that the signal is unique
and significant in the RV BFPs shown in panel P31, P32 and P33 of
Fig. \ref{fig:BFP_GJ229}, it is likely related to a planet. The moving
periodogram shown in Fig. \ref{fig:MP_GJ229c} supports a reasonable
time consistency and displays higher significance in the recent high-cadence HARPS data.   

The long period signal at a period of about 50,000 days
corresponds to GJ 229B, which is a brown dwarf detected through direct
imaging \citep{nakajima95,golimowski98}. Thanks to a time span of
nearly 20 years of combined data, we exclude an orbital period of less
than 15,800 days with a false alarm probability of 1\%. The minimum mass
of GJ 229B is about 515\,$M_\oplus$ or 1.62\,$M_{\rm Jup}$, which is
much lower than the estimation of $\sim$20\,$M_{\rm Jup}$ by
\cite{nakajima95} based on cooling models of brown dwarfs and the
  recent estimation of 72\,$M_{\rm Jup}$ by \cite{brandt19} based on a
combined analysis of KECK RVs and astrometric data. This suggests a
relatively face-on orientation for the system, consistent with an
inclination of $13_{-12}^{+10}$\,deg estimated by \cite{brandt19}.
Since the orbital period of GJ 229B is much longer than the RV timespan, we do not show the moving periodogram for this signal.
In Fig. \ref{fig:BFP_GJ229}, we see a significant activity signal at a period of 278 days in P49, P50, P51, P52, P53, P61, P63 and P64. It is probably caused by stellar rotation or magnetic cycles.

\item \textbf{GJ 422 (HIP 55042)} A signal at a period of 20.129 days
  with a minimum mass of 10.4\,$M_\oplus$ is identified in the
  combined UVES, HARPSpre and HARPSpost data. This signal is not found in
  activity indices (see Fig. \ref{fig:BFP_GJ422}). It is found to
  be strong in all of the three sets (P4 to P12) as well as in the combined
  set (P1 to P3). A signal around 8 days is found to have strong powers in the BFPs for the H$\alpha$ of HARPSpre (P35) and for other indices (P31 and P37). The 26-day signal reported by T14 is not as significant as this signal, as seen
  in Fig. \ref{fig:BFP_GJ422}. Considering that these two signals
  are probably related, we still uses GJ 422 b to name this candidate.

\item \textbf{GJ 433 (HIP 56528)} There are three signals with periods
  of 7.37, 36.1 and 5000 days and minium masses of 5.94, 4.94,
  28.8\,$M_\oplus$ found in the combined HARPS, KECK, PFSpre, PFSpost
  and UVES data (see Fig. \ref{fig:BFP_GJ433}). The 7.37-day signal is
  found by T14. This signal is unique and consistent over time, as
  shown by the moving periodogram in Fig. \ref{fig:MP_GJ433b}. The
  36-day signal is significant in the HARPS set and shows high ln(BF)
  in the BFPs for KECK, PFS and UVES. The moving periodogram shown in
  Fig. \ref{fig:MP_GJ433d} demonstrates that this signal is unique and consistent over time. 

T14 find a long period signal with an optimal period of 2950 days
although the period is consistent with any values larger than 1900
days. In Fig. \ref{fig:BFP_GJ433}, we see strong power around 2947.6
days in the BFP for the HARPS data when the 7.37-day and 36-day signals (P34-36) have been subtracted. However, the signal does not appear in the BFPs for the residuals of PFSpre (P40-P42) and UVES (P43-P45). 
With more data sets spanning a longer period (see the phase curve
for GJ 433 d in Fig. \ref{fig:phase}), we are able to constrain the
period to be 5093$\pm$610 days. In the BFPs for the combined
residuals, the power around 5000 days is high in the BFP for white
noise model (P31) and is a bit lower in the BFPs for MA and AR models
(P32-P33). This effect for long period signals is expected
because a linear trend is fit to the data in the calculation of
BFP. Thus we rely on a full MCMC posterior sampling to identify and
constrain the signal, leading to ${\rm ln(BF_5)}=10.5$ for the 3-planet
model compared with the 2-planet model. Since the orbital period of
this signal is longer than the RV timespan, the moving periodogram is
not able to demonstrate time consistency. This signal corresponds to
super-Neptune with a semi-major axis of about 4.7\,au, equivalent to a
separation of 0.5\,as to GJ 433. 

As Fig. \ref{fig:BFP_GJ433} shows, the activity signal at a period
of 107.3 days is significant in the BFPs for the FWHM and NaD2 indices
from HARPS. Thus we consider this signal as the rotation period,
differing from 73.2$\pm$16.0\,d inferred by \cite{suarez15} using the
Ca II H\&K and H-alpha indicators.

\item \textbf{GJ 620 (HIP 80268)} Two signals with periods of 7.65 and
  27.2 days and minium masses of 7.38 and 10.08\,$M_\oplus$ are
  identified in the combined UVES and HARPS data. This signal is not
  found in the activity indices (see Fig. \ref{fig:BFP_GJ620}). Considering there are only 5 UVES and 23 HARPS RV data points, we use one time window to cover the former and one window to cover the latter to calculate the moving periodogram. In Fig. \ref{fig:MP_GJ620b}, the power around 7.65 days is consistent over time but is not unique for the UVES data due to the poor sampling of the orbital phase. Similarly, the 27.2-day signal shows time consistency in the periodogram shown in Fig. \ref{fig:MP_GJ620c}.
  
\item \textbf{GJ 739 (HIP 93206)} Two signals with periods of 45.3
  and 266 days and minimum masses of 8.49 and 47.52\,$M_\oplus$ are
  identified in the combined UVES and HARPS data.
  These signals are not found in activity indices (see
  Fig. \ref{fig:BFP_GJ739}). In Fig. \ref{fig:MP_GJ739b}, the moving
  periodogram shows time consistency for these signals
  although the signal is not unique in the recent HARPS data due to
  poor sampling. The moving periodogram shown in
  Fig. \ref{fig:MP_GJ739c} demonstrates the time consistency for
  the 266-day signal. 

  \item \textbf{GJ 911 (HIP 117886)} A signal at a period of 2.79
    days and with a minium mass of 8.21\,$M_\oplus$ is identified in
    the combined UVES, HARPS and KECK data. As seen
    from Fig. \ref{fig:BFP_GJ911}, this signal is not found in activity indices (see Fig. \ref{fig:BFP_GJ911}). This signal is significant in the UVES
    set (P7 and P9) and corresponding $\ln{\rm BF}_5=6.23$
    suggests high statistical significance although it is not found to be
    significant in the BFPs for the combined data set (P1, P2 and
    P3). To visualize the time consistency of this signal, we
    create the moving periodogram by calculating the BFP for the RV
    points in a time window. The window moves in 10 steps to cover the
    whole RV timespan. The BFPs are normalized such that the maximum
    lnBF difference for a time window is one. The size of the window
    is automatically determined such that it is large enough to allow
    a minimum number of 10\% of the total number of RV points in each
    of the 10 windows (see Paper I for details). We show the moving
    periodogram for GJ 911 in Fig. \ref{fig:MP_GJ911b}. Although the
    normalized ln(BF) around 2.79 days is high for the recent KECK
    measurements  ($>$2500 days after the first epoch), there are
    multiple signals with similar significance. The power around 2.79 days for the previous UVES RVs is strongly evident despite having less contrast
 compared with other periods. Thus the
    overall evidence favors a time consistent signal at a period of
    2.79 days. The other signals are either not as significant or not
    as consistent as this signal. 
  \item \textbf{GJ 3082 (HIP 5812)} A signal around 11.9-day with a
    minium mass of 8.77\,$M_\oplus$ is identified in the combined
    HARPS and UVES data. In Fig. \ref{fig:BFP_GJ3082}, the BFPs for the data show high and
    unique power at this period (P1, P2 and P3). Compared with the
    white noise BFP, the red noise BFPs show more unique and
    significant power for this signal. We also identify a significant
    activity signal at a period of 123 days in the BFPs for FWHM and
    NaD2 of HARPS. We show the moving periodogram for this signal in
    Fig. \ref{fig:MP_GJ3082b}. This signal is unique and
    consistent over time although multiple aliases are shown in the
    BFP for the recent HARPS RVs. 
\end{itemize}

\subsection{Comments on previous planet candidates} \label{sec:T14}
Based on our combined analysis of the other UVES targets, we comment
on the planetary candidates reported in table 4 of T14 as follows. 
\begin{itemize}
  \item \textbf{GJ 27.1 b} It is likely due to activity. The signals at periods of 31.8 and 10.3 days are identified in the combined UVES, KECK and HARPS data. We show them in Fig.
  \ref{fig:BFP_GJ27.1} together with the 15.819-day signal reported by
  T14. The 31.8-day signal is found to be significant in the activity indices
  of BIS, FWHM, RHK, H$\alpha$ and NaD1 of the HARPS set, indicating a
  activity origin. The 15.8-day signal reported by T14 is nearly
  half of the activity signal and strong powers around this period are
  seen in the BFPs for FWHM (P14), R$_{\rm HK}$ (P15) and
  H$\alpha$ (P16) of HARPS. The 10.3-day signal is not as significant
  as the 31.8-day signal in the BFPs for activity indices although we
  still find strong powers around similar periods in the BFPs for FWHM
  (P14) and NaD2 (P18) of HARPS. Considering that the 31.8-day is significant and unique enough to
  disguise itself as a Keplerian signal, a comprehensive diagnosis of
  activity indices are necessary for reliable planet detections.
  
\item \textbf{GJ 160.2 b} It is not identified. We do not find this signal through MCMC
  sampling of the posterior. As seen in Fig. \ref{fig:BFP_GJ160.2},
  no signficant and unique signals are found in the red noise BFPs for the
  combined data set (P2 and P3) despite strong powers in the white
  noise BFPs (P1). Since the 5.2-day signal found by T14 is not
  visible in the new UVES set or in other data sets, it is unlikely to
  be Keplerian and might arise from data reduction. 
  
\item \textbf{GJ 682 b and c} They are probably due to activity. As seen in Fig. \ref{fig:BFP_GJ682}, we
  did not find these two signals with periods of 17.48 and 57.32 days
  in the BFPs nor in the MCMC posterior samples. The signals with strong powers in the white noise BFPs (P1) may be caused by correlated noise since we do not see them in red noise BFPs (P2 and P3). Moreover, we find the 17.48-day signal is
  found to be significant in the BFP for BIS of HARPS, indicating an
  activity origin. In the BFP for FWHM of HARPS, we find another
  signal around 6.87 days to be significant. There is also strong
  power around this period in the BFP for R$_{\rm HK}$ of HARPS. Thus
  this signal might be caused by the differential rotation of the star. 
\end{itemize}

\input{fig}

\subsection{Dynamical stability of GJ 180 and GJ 229} \label{sec:stability}
The angular momentum deficit (AMD) is typically used to study the
  dynamical stability of planetary systems
  \citep{laskar00,laskar17}. Although the AMD is efficient in
  generally assessing stability, AMD-unstable systems can be
  stablized through mechanisms such as mean motion resonances
  \citep{laskar17}. To avoid such caveats, we examine the orbital stability of the GJ 180 and GJ 229 systems with N-body integrations using the Mercury integration package \citep{chambers99}.

\begin{figure*}
  \centering
  \includegraphics[scale=0.5]{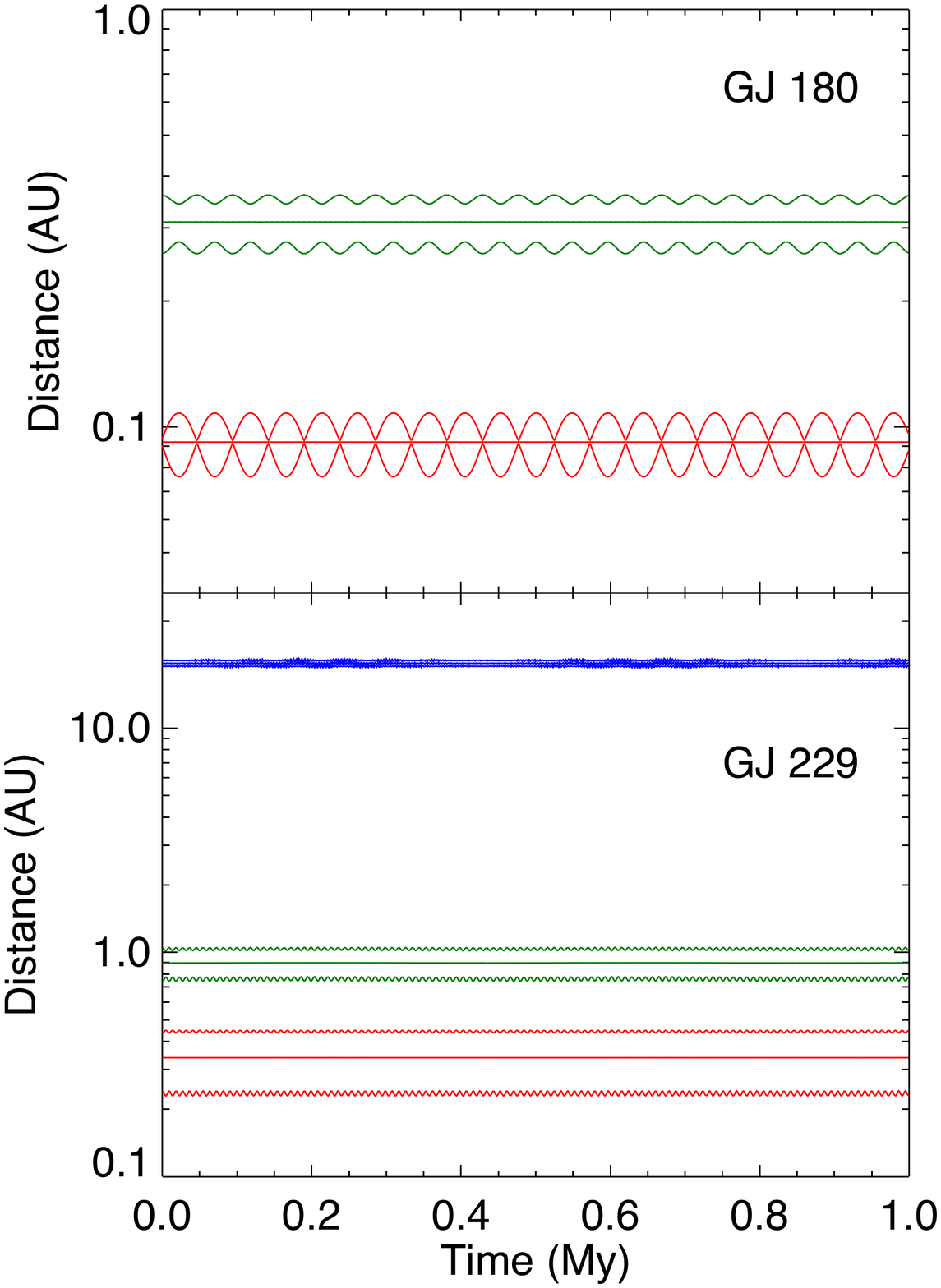}
  \caption{One sample of the dynamical simulations for GJ 180 (upper)
    and GJ 229 (lower) over one million years. The three curves of
    each color show the apoastron distance, semi-major axis and
    periastron distance for a planet. In the upper panel, the red
    curves represent GJ 180 b and the green curves represent GJ 180
    d. In the lower panel, the red, green, and blue curves represent
    GJ 229A c, GJ 229 b and GJ 229B, respectively.}
  \label{fig:stability}
\end{figure*}

For the 2-planet GJ 180 system, we examine a grid of initial
eccentricities $e$ and masses $m_p$ for GJ 180 d spanning 1 standard
deviation above and below the mean values listed in Table
\ref{tab:planet}. The other elements for GJ 180 d and the mass and
elements for GJ 180 b were held at their mean values. For each grid of
$m_p$ and $e$ values, we consider 3 inclinations for the system as a
whole: 30, 60 and 90 degrees. The masses for both planets were
adjusted accordingly while keeping the planetary orbits coplanar. The
integrations used a second-order mixed-variable symplectic (MVS)
integrator with a step size of 0.7 days. The upper panel of the
Fig. \ref{fig:stability} shows a typical case with $I=90$ degrees and
using the nominal mass and eccentricity for GJ 180 d. In this case, and all the other cases we considered, there is no indication of instability during the 1\,My integration. The semi-major axes are almost constant, while the eccentricities undergo small, periodic oscillations. The orbits never approach each other. Although these integrations do not prove the system is stable, they clearly suggest that this is the case.

For the GJ 229 system, we included planets GJ 229A c and GJ 229A b as
well as the likely brown dwarf GJ 229B. We examined a grid of values
for e and M of GJ 229A c spanning one standard deviation above and
below the mean values given in Table \ref{tab:planet}, holding all
other quantities fixed at their nominal values. For each grid, we
considered two inclinations for the system such that the masses of
each object were either 10 or 20 times the nominal values for an
edge-on system (corresponding to inclinations of about 5.7 and 2.9
degrees respectively). These choices reflect the fact that GJ 229B is
probably a brown dwarf, which suggests that the system is nearly face
on to us. The orbits of the planets and brown dwarf were assumed to be
coplanar. The integrations used an MVS integrator with a step size of
4 days. The lower panel of the Fig. \ref{fig:stability} shows a
typical case with $I=5.7$ degrees and the nominal mass and
eccentricity for GJ 229A c. As with GJ 180, the orbits show no sign of
instability over 1\,My, and this was true for all the cases we considered here.

\section{Conclusion}\label{sec:conclusion}
We identify sixteen planet candidates orbiting nine nearby M dwarfs
from the combined analysis of the UVES, HIRES/KECK, HARPS and PFS velocity
data sets. Among these candidates, GJ 173 b, GJ 180 d, GJ 229A c, GJ 422 b, GJ 433 d and GJ 3082 b are five new planets corresponding to RV signals that are statistically significant, robust to the choice
of noise model, unique and consistent over time. Thus they are likely
real planets. We confirm three previously reported planet
candidates and find eight planet candidates which need additional observations to be further confirmed.

In the new sample of candidates, two temperate super-Earths, GJ 180 d and
GJ 229A c, are located in the habitable zones of their host stars. Due to
their proximity, the separation between the host stars and the potentially
habitable planets is in these systems is 25\,mas and 59\,mas respectively.
These are good candidates for future direct imaging by future facilities. 

GJ 433 c is a cold super-Neptune candidate located in a mass-period regime rarely
explored. It is separated from its host star by about 0.5\,as and thus
is a good candidate for direct imaging. A comprehensive RV survey of these planets would
help us better understand Neptune-like planets. 

Among the sixteen candidates, four planets have been reported by
T14 and GJ 229B, a T-type brown dwarf, has been discovered by
\cite{nakajima95} through direct imaging. With updated UVES and
HARPS data combined with KECK and PFS, we provide improved 
constraints on the orbits of these planets. However, we fail to
confirm GJ 27.1 b, GJ 160.2 b, GJ682 b and c reported by T14
probably due to our use of updated UVES data and more RV sets. Some of these signals probably have an activity
origin through our diagnosis of the periodograms for activity indices. 

There are five multiple-planet systems. GJ 180, GJ 620, and GJ 739
each hosts two planets. GJ 229 and GJ 433 each hosts three
planets\footnote{Here we count GJ 229 B as a planet.}. These planets
are likely stable because their orbits are not eccentric and are well
separated. In particular, we examine the stability of GJ 229 and GJ 180
systems by simulating the orbits of their planets over one million
years. We find that all planets in these two systems are stable,
suggesting long-term stable orbits of the two temperate planets GJ
229A c and GJ 180 d. 

Our detection of these Keplerian signals demonstrates the
value of a combined analysis of multiple RV data sets to extend the
time baseline, and avoid instrumental bias.  We also highlight the
importance of the moving periodogram in the confirmation of
signals. The time-consistent uniqueness and significance of GJ 173 b
(Fig. \ref{fig:MP_GJ173b}), GJ 180 d (Fig. \ref{fig:MP_GJ180c}), GJ 229A c
(Fig. \ref{fig:MP_GJ229b}), GJ 433 b (Fig. \ref{fig:MP_GJ433b})
and GJ 433 d (Fig. \ref{fig:MP_GJ433d}) are well presented by
the moving periodograms. Hence we recommend moving periodograms as a tool for signal visualization. Our identification of false positives
through BFPs of activity indices demonstrate an essential role of
comprehensive periodogram analysis in planet discoveries. Our work is
based on an updated data reduction of the UVES data collected in
2009. This demonstrates the on-going necessity of better analysis of archived RV data to reveal small signals. Since we can not go back in time, this data represents our earliest epoch for future discoveries.


\section*{Acknowledgements}
This work has made use of data from the European Space Agency (ESA)
mission {\it Gaia} (https://www.cosmos.esa.int/gaia), processed by the
{\it Gaia} Data Processing and Analysis Consortium (DPAC,
https://www.cosmos.esa.int/web/gaia/dpac/consortium). Funding for the
DPAC has been provided by national institutions, in particular the
institutions participating in the {\it Gaia} Multilateral
Agreement. This research has also made use of the Keck Observatory Archive
(KOA), which is operated by the W. M. Keck Observatory and the
NASA Exoplanet Science Institute (NExScI), under contract with
the National Aeronautics and Space Administration.  This
research has also made use of the services of the ESO Science
Archive Facility, NASA's Astrophysics Data System Bibliographic
Service, and the SIMBAD database, operated at CDS, Strasbourg, France.
Support for this work was provided by NASA through Hubble
Fellowship grant HST-HF2-51399.001 awarded by the Space Telescope
Science Institute, which is operated by the Association of
Universities for Research in Astronomy, Inc., for NASA, under contract
NAS5-26555. The authors acknowledge the years of technical support from LCO staff in the successful operation of PFS, enabling the collection of the data presented in this paper.
\software{R package magicaxis \citep{robotham16},
fields\citep{fields}, minpack.lm \citep{elzhov16},
  nortest\footnote{\url{https://cran.r-project.org/web/packages/nortest/index.html}},
  aTSA\footnote{\url{https://cran.r-project.org/web/packages/aTSA/index.html}}}

\appendix
\restartappendixnumbering 
\section{Tests for priors}\label{sec:prior}
To test the sensitivity of parameter inference and model
  selection to eccentricity priors, we compare the parameter values
  and ln(BF) for the semi-Gaussian eccentricity prior with standard
  deviation of 0.2 (used in this work) and 0.4 in Table
  \ref{tab:prior}.
\begin{table}
\caption{Sensitivity of paramerer inference and model selection to
 eccentricity priors. The optimal values of parameters corresponds to
 the MAP solution and their undertainties are determined at the 10 and
 90\% quantiles of the MCMC posterior samples.}
\label{tab:prior}
\centering
\begin{tabular}{c|*{3}{c}|*{3}{c}}
\hline
  Targets&\multicolumn{3}{c|}{$\mathcal{N}(0,0.2)$}&\multicolumn{3}{c}{$\mathcal{N}(0,0.4)$}\\
         &$P$ (day)&$K$ (m/s)&$e$&$P$ (day)&$K$ (m/s)&$e$\\\hline
  GJ 180 d &$106.3_{-0.2}^{+0.13}$ &$2.1_{-0.29}^{+0.40}$ &$0.16_{-0.055}^{+0.05}$ &$106.4_{-0.18}^{+0.14}$ &$2.9_{-1.0}^{+0.0}$ &$0.26_{-0.22}^{+0.0007}$\\
  GJ 229A c&$122.0_{-0.25}^{+0.16}$ &$1.9_{-0.13}^{+0.53}$ &$0.23_{-0.14}^{+0.074}$ &$121.9_{-0.01}^{+0.32}$ &$2.3_{-0.73}^{+0.00}$ &$0.32_{-0.17}^{+0.10}$ \\
  GJ 422 b&$20.13_{-0.0068}^{+0.0069}$ &$4.2_{-0.12}^{+0.78}$ &$0.10_{-0.059}^{+0.063}$&$20.13_{-0.0059}^{+0.0076}$ &$4.3_{-0.54}^{+0.55}$ &$0.10_{-0.030}^{+0.14}$\\
  GJ 433 d&$36.05_{-0.014}^{+0.028}$ &$1.4_{-0.22}^{+0.39}$ &$0.028_{-0.011}^{+0.11}$&$36.07_{-0.035}^{+0.012}$ &$1.6_{-0.46}^{+0.085}$ &$0.20_{-0.15}^{+0.16}$\\
  GJ 3082 b&$11.94_{-0.0037}^{+0.061}$ &$4.2_{-1.2}^{+0.69}$ &$0.26_{-0.18}^{+0.11}$ &$12.00_{-0.063}^{+0.0061}$ &$3.9_{-1.1}^{+0.76}$ &$0.40_{-0.26}^{+0.11}$ \\\hline
  \end{tabular}
\end{table}
  
\section{Tests for residuals}\label{sec:residual}
We test the normality, autocorrelation and stationarity of residuals
using the Anderson-Darling test \citep{anderson52}, the Ljung-Box test
\citep{box70,ljung78}, and the Augmented Dickey-Fuller
\citep{dickey79}, respectively. Augmented Dickey-Fuller (DF)
tests of various types with lags of less than four show a p-value of
$\le$0.01. The DF tests for larger time lags show slightly higher p-value though
it is still less than 0.05. However, this low p-value may not suggest
non-stationarity because DF is typically applied to regularly spaced
time series. To test this, we fit a linear trend to each residuals and
apply the DF test to the new residuals without any lag (i.e., traditional DF test). The p-value is still $\le$0.01 and thus suggests a linear trend
although the best-fit linear trend is subtracted from the residual if
there is any. Hence, the DF test is not appropriate for testing the
stationarity of irregularly spaced time series.

We report the statistic and p-value for the Anderson-Darling (AD) and
Ljung-Box (LB) tests for each target in Table \ref{tab:tests}. The AD
normality tests suggest that the RV residual for GJ 173 does not
follow a Gaussian distribution, suggesting potential signals in the residuals. This is
consistent with the two strong signals around periods of 20 and 60
days shown in the residual BFPs (see P13-P15 of
Fig. \ref{fig:BFP_GJ173}). The AD tests for the other residuals show
also relatively low p-value due to potential signals in RV
residuals. Since the moving average model used in this study is
stochastic and cannot model insignificant periodic signals, we expect
considerable time-correlated noise or signal over long timescales in
RV residuals.

The p-values of LB tests for GJ 180, GJ 229A, GJ 422, GJ 433, GJ 620,
GJ 739 and GJ 911 are high, suggesting insignificant autocorrelation
in residuals. However, the LB tests for GJ 173 and GJ 3082 have low
p-values, indicating considerable autocorrelation. For these two
targets, we model the noise using white noise models according to the
Bayesian model selection scheme introduced in section
\ref{sec:posterior}. Thus, in the RV residuals, we expect to find
time-correlated noise, which is not significant enough to be modeled
by the MA model. 
\begin{table}
\caption{Statistic and p-value of the Anderson-Darling,
  Ljung-Box (LB) tests for the residuals.}
\label{tab:tests}
\centering
\begin{tabular}{*{5}{c}}
\hline\hline
Target&AD statistic&AD p-value&LB statistic&LB p-value\\\hline
  GJ 173 &1.3 & 0.0027 & 4.5 & 0.033 \\
  GJ 180 & 0.52 & 0.18 & 0.98 & 0.32 \\
  GJ 229A& 3.1 & $8.6\times 10^{-8}$ & 0.93 & 0.33 \\
  GJ 422 & 2.4 & $3.5\times 10^{-6}$ & 0.34 & 0.56 \\
  GJ 433 & 0.46 & 0.26 & 1.6 & 0.21 \\
  GJ 620 & 0.67 & 0.073 & 0.68 & 0.41 \\
  GJ 739 & 0.86 & 0.026 & 0.32 & 0.57 \\
  GJ 911 & 1.9 & $8\times 10^{-5}$ & 0.0072 & 0.93 \\
  GJ 3082 & 0.2 & 0.88 & 6.3 & 0.012\\\hline                               
  \end{tabular}
\end{table}
\bibliographystyle{aasjournal}
\bibliography{nm}  
\end{document}

%% file: fig.tex
\begin{figure*}
  \centering
  \includegraphics[scale=0.5]{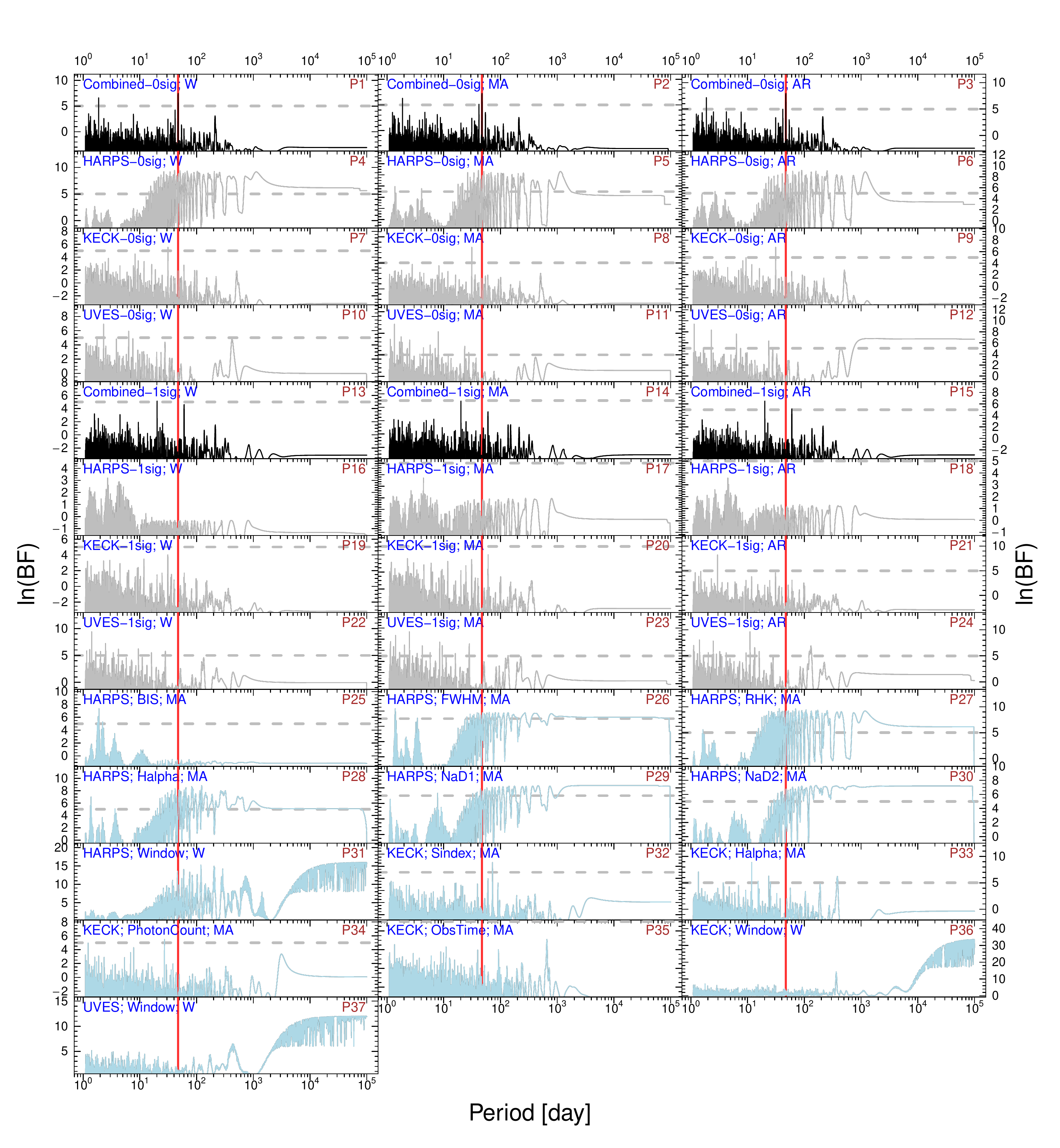}
  \caption{BFPs for GJ 173. The black BFPs are calculated for the
    combined data subtracted by signals subsequently while the grey
    BFPs are for individual data sets. The lightblue BFPs are
    for activity indices and window functions. The top left legend in
    each  panel denotes the name of RV data set or activity index and
    the number of signals subtracted from the data. The top right
    legend in each panel shows the panel number for reference. The red
    lines denote the planetary signal at a period of 47.3 days. The horizontal dashed lines show the threshold of ln(BF)=5.}
  \label{fig:BFP_GJ173}
\end{figure*}

\begin{figure*}
  \centering
  \includegraphics[scale=0.4]{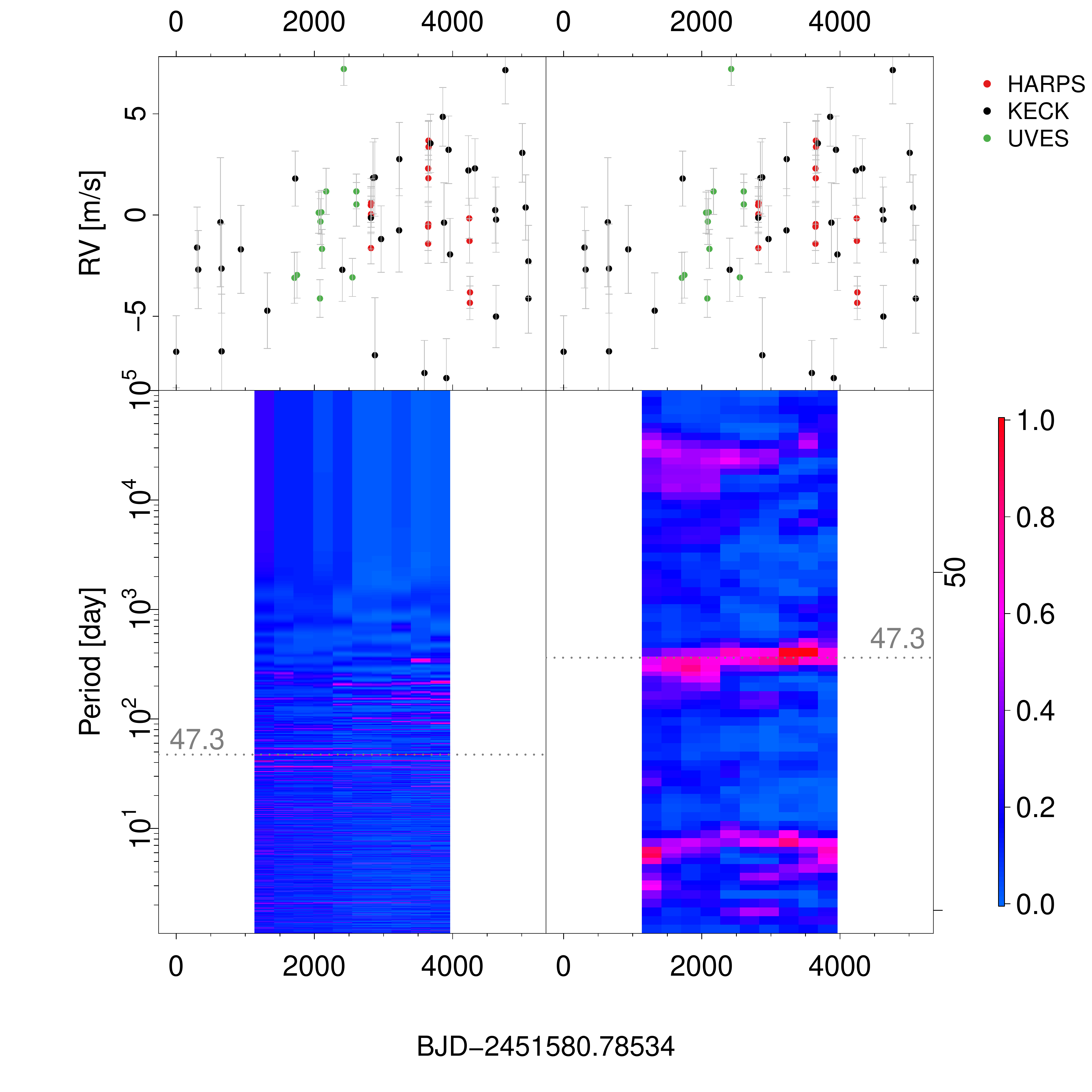}
   \caption{Moving periodogram for GJ 173. The top two panels show the same combined RV data subtracted by the best-fit MA red noise. The bottom left panel shows the BFPs as a function of period and time window with a size of about 2000 days (double of the width of the gap between the left y axis and the left edge of the color-coded BFPs). The bottom right panel is a zoom-in of the bottom left panel to optimize the visualization of the signal. The annual aliases of the 47.3-day signal at periods of 54 and 42 days are visible and coded by red color. }
  \label{fig:MP_GJ173b}
\end{figure*}

\begin{figure*}
    \vspace{-0.3in}
   \hspace{-0.5in}\includegraphics[scale=0.5]{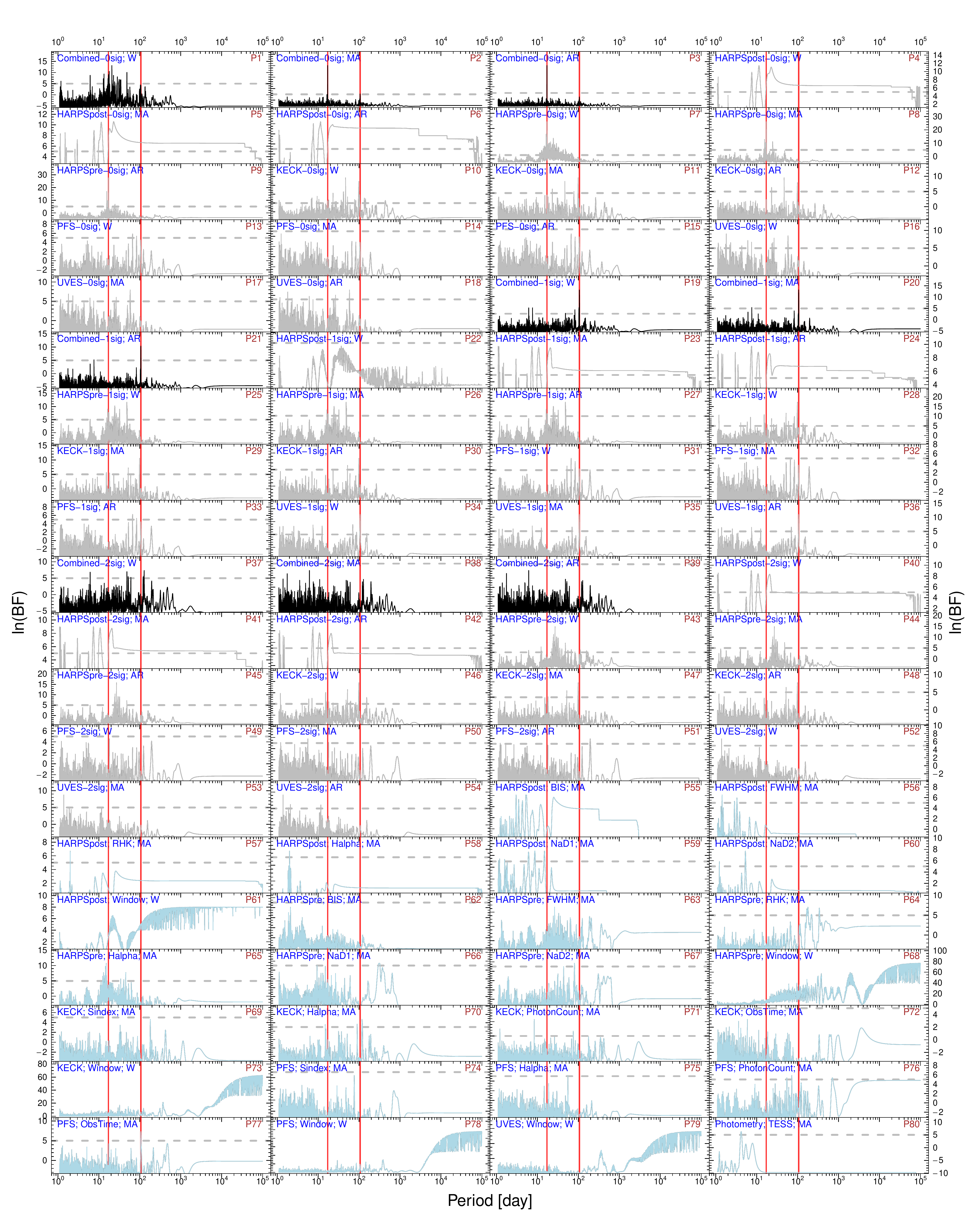}
    \vspace{-0.3in}
  \caption{BFP for GJ 180. The red lines denote the signal at periods of 17 and 106 days, which are subtracted from the raw RV data subsequently to calculate residual BFPs. The P80 panel shows the BFP for the TESS photometry data \citep{ricker14}. }
  \label{fig:BFP_GJ180}
\end{figure*}

\begin{figure*}
  \centering
  \includegraphics[scale=0.4]{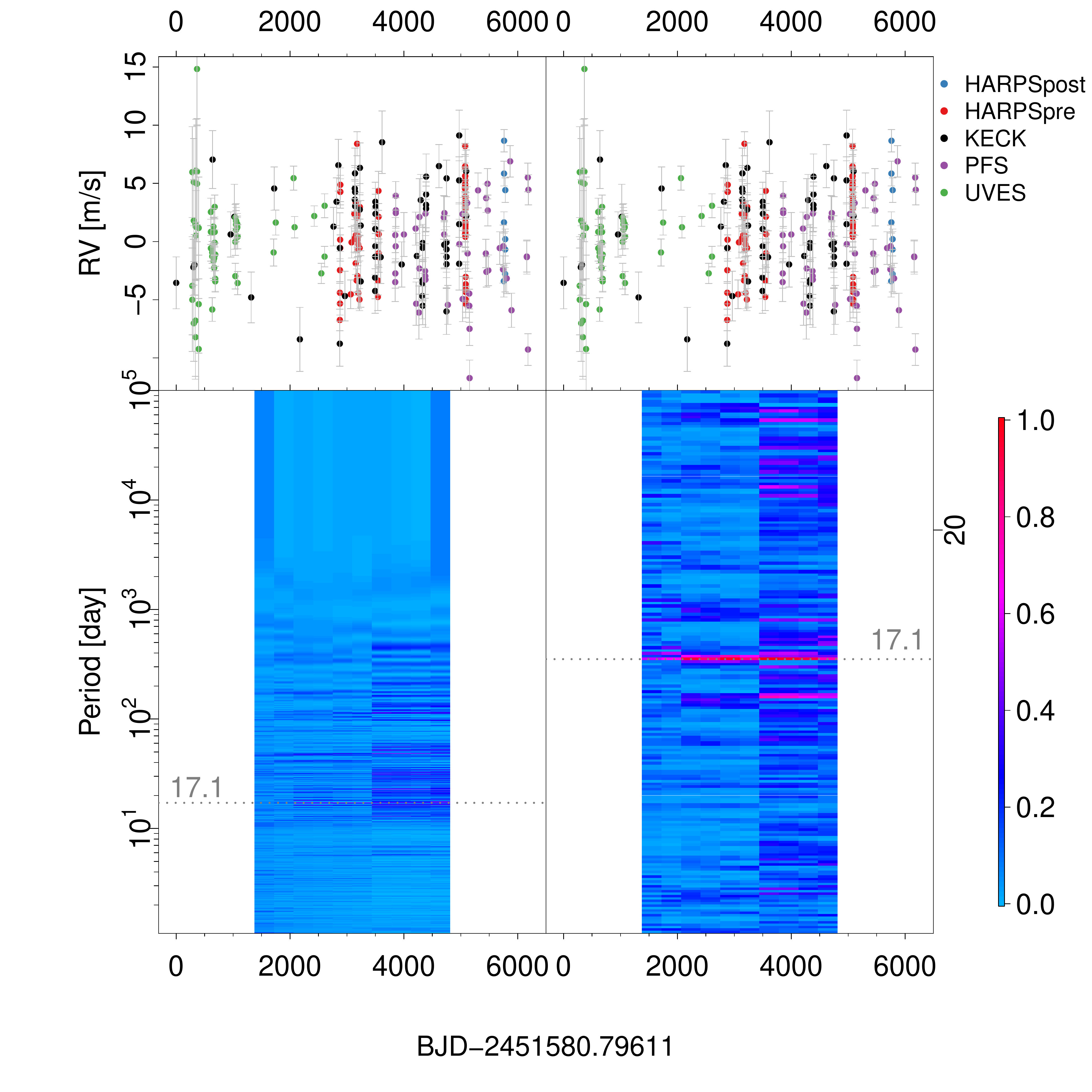}
   \caption{Moving periodogram for GJ 180 b. The annual aliases of the
     17.9-day signal at periods of 16.3 and 17.9 days are also visible.}
  \label{fig:MP_GJ180b}
\end{figure*}

\begin{figure*}
 \centering
  \includegraphics[scale=0.4]{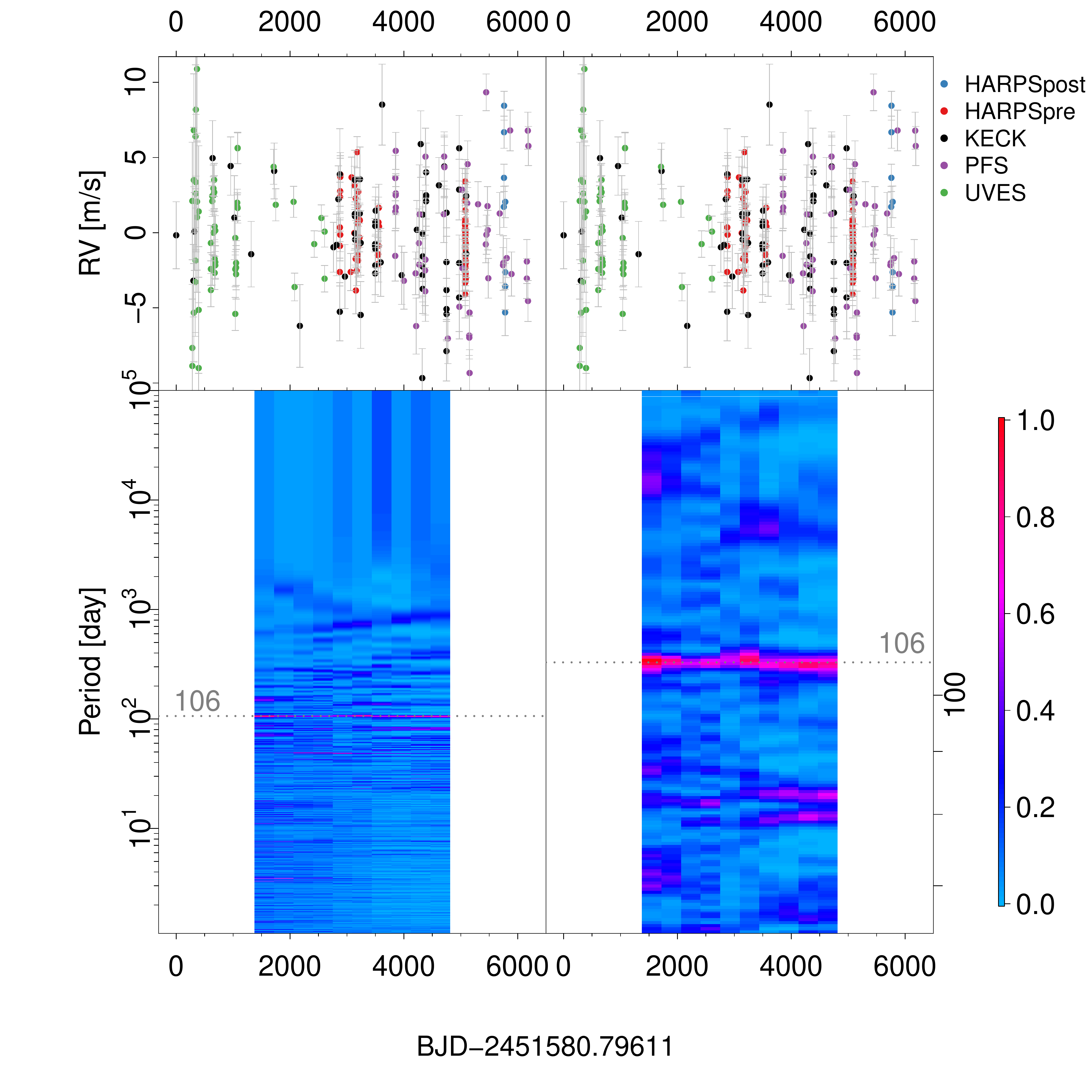}
   \caption{Moving periodogram for GJ 180 c. The annual aliases of the
     106-day signal at periods of 82 and 149 days are visible though not
     consistent over time.}
  \label{fig:MP_GJ180c}
\end{figure*}

\begin{figure*}
  \vspace{-0.3in}
  \hspace{-0.2in}\includegraphics[scale=0.46]{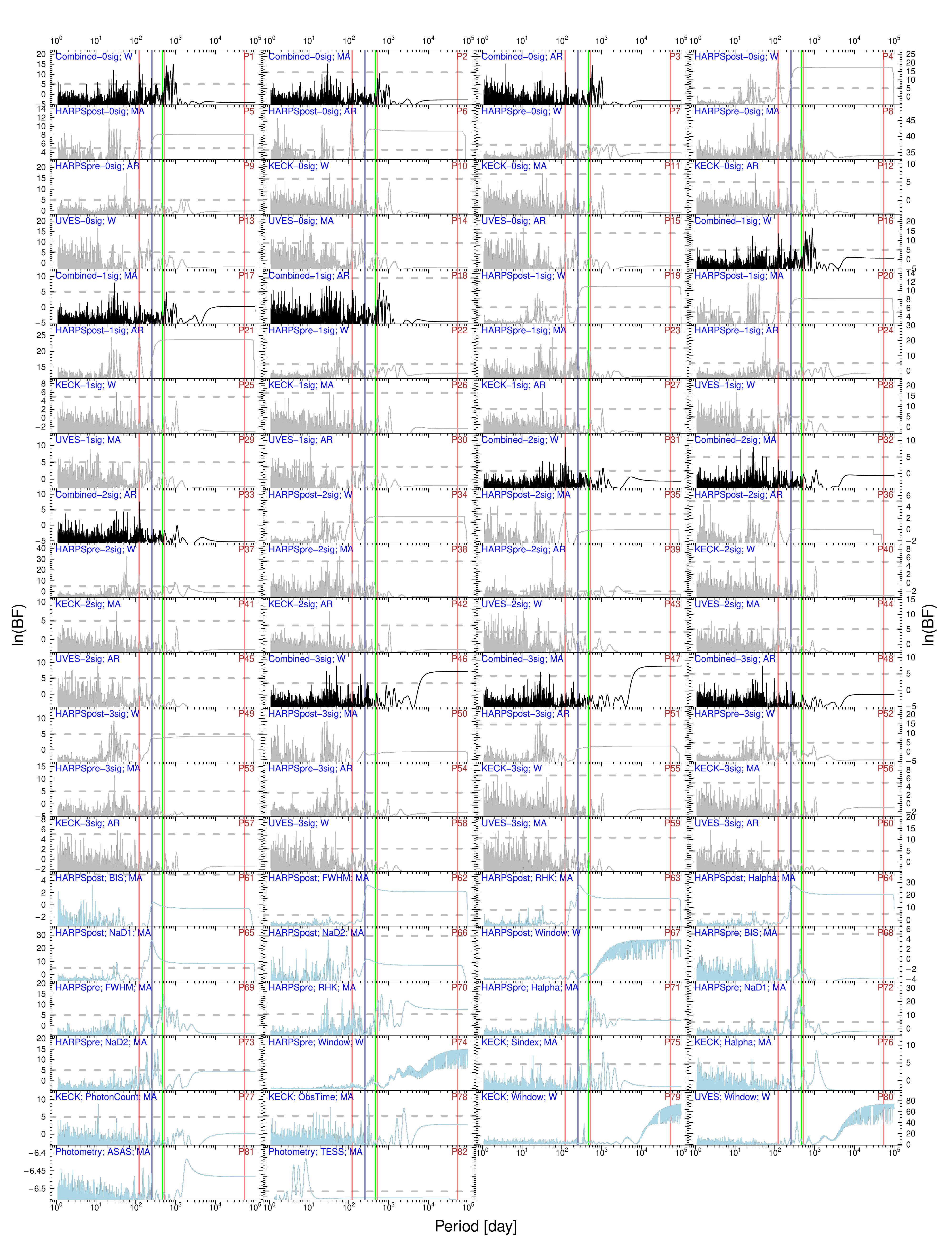}
  \vspace{-0.2in}
  \caption{BFP for GJ 229A. The red lines denote signal periods of
    52900, 122 and 520 days, which are subtracted from the raw RV data
    subsequently to calculate the residual BFPs. The darkblue lines denote the activity signal with a period of 278 days. The cyan line
    denotes the 471-day signal reported by T14. The BFPs for
    individual data sets subtracted by three signals are not shown to
    simplify the figure.}
  \label{fig:BFP_GJ229}
\end{figure*}

\begin{figure*}
  \centering
  \includegraphics[scale=0.4]{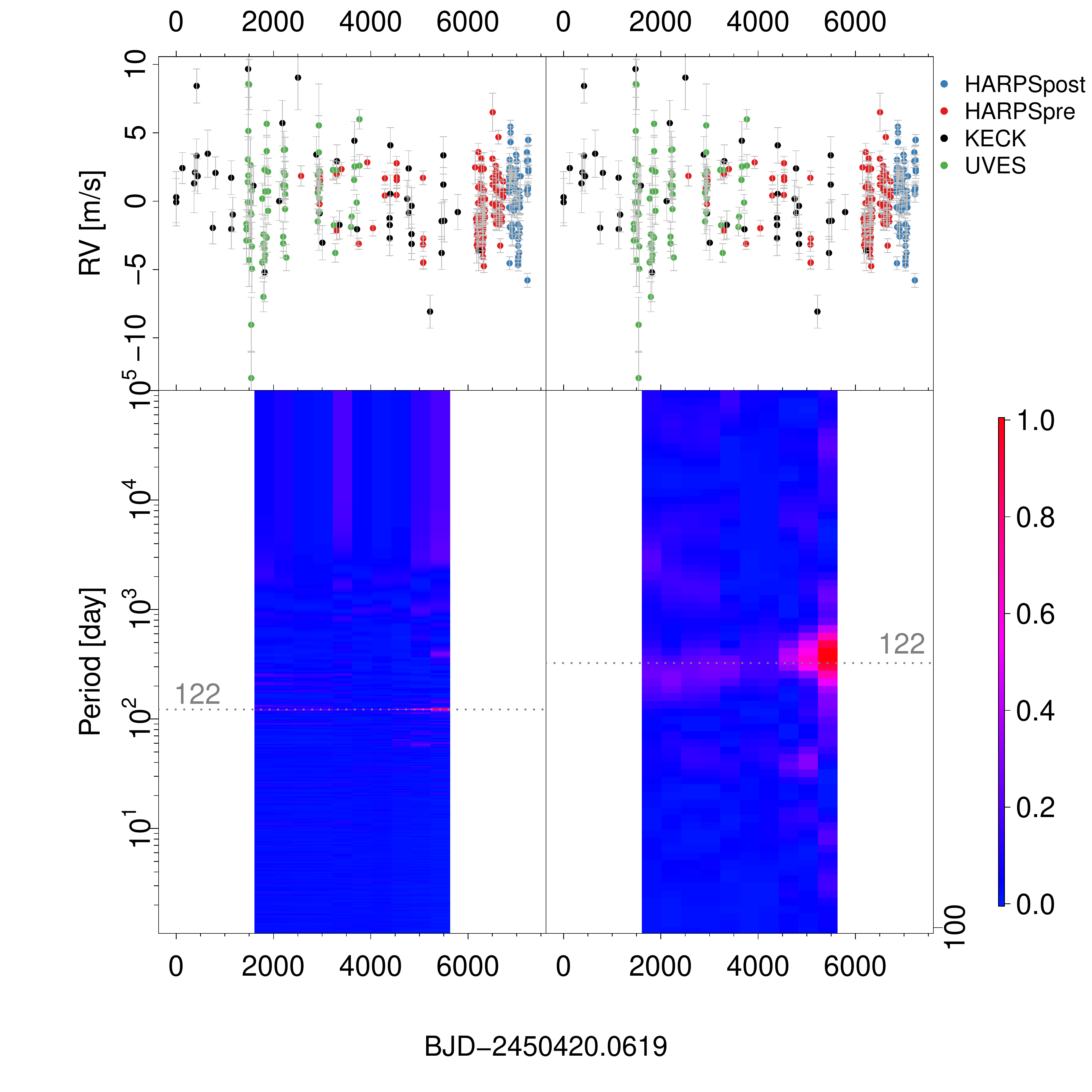}
   \caption{Moving periodogram for GJ 229A c. The 122-day signal is most significant in the HARPSpost data set due to its high cadence.}
   \label{fig:MP_GJ229b}
   \end{figure*}

   \begin{figure*}
  \centering
  \includegraphics[scale=0.4]{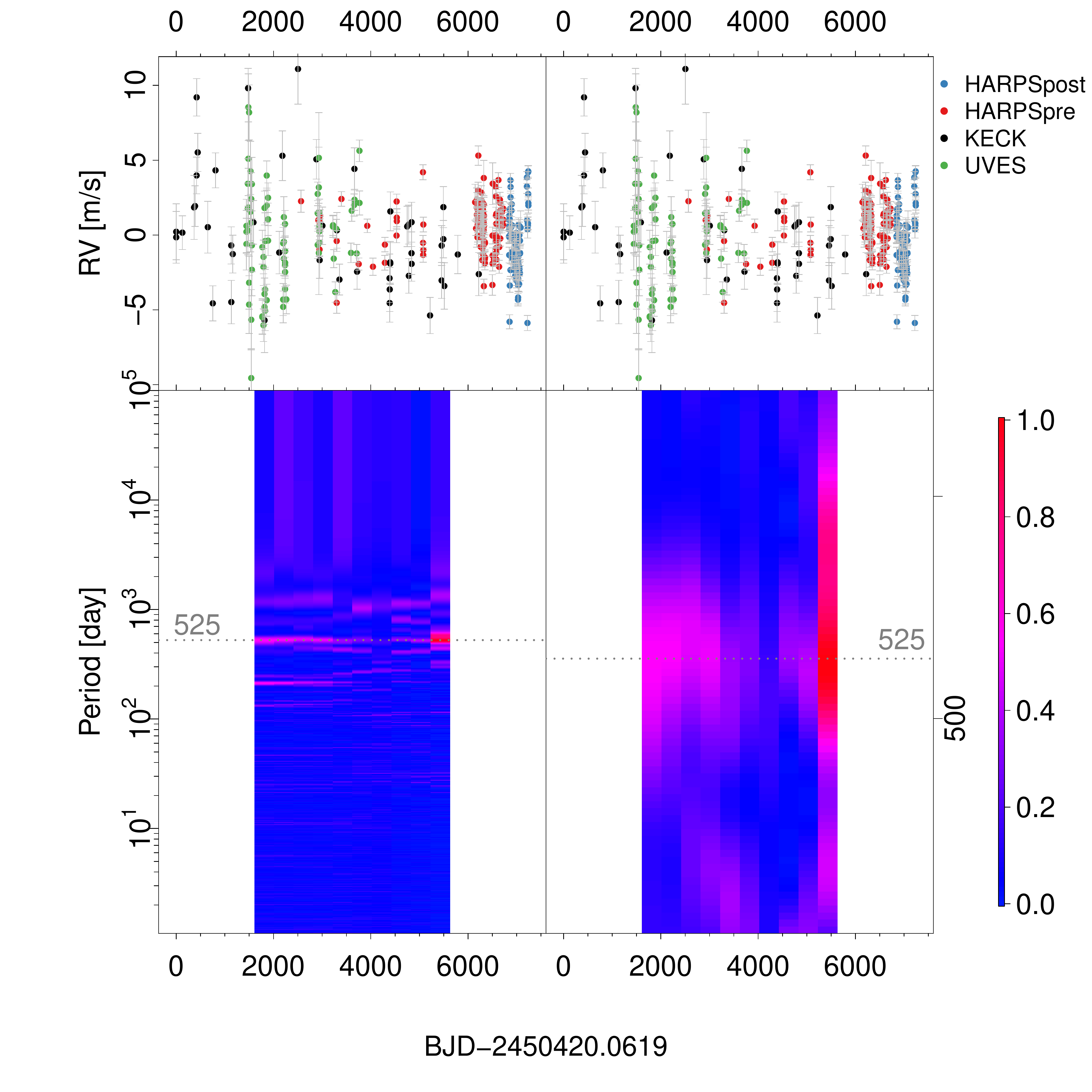}
   \caption{Moving periodogram for GJ 229A b. The annual aliases of the
     525-day signal at periods of 215 and 1200 days are visible though not
     consistent over time.}
  \label{fig:MP_GJ229c}
\end{figure*}

\begin{figure*}
  \centering
  \includegraphics[scale=0.5]{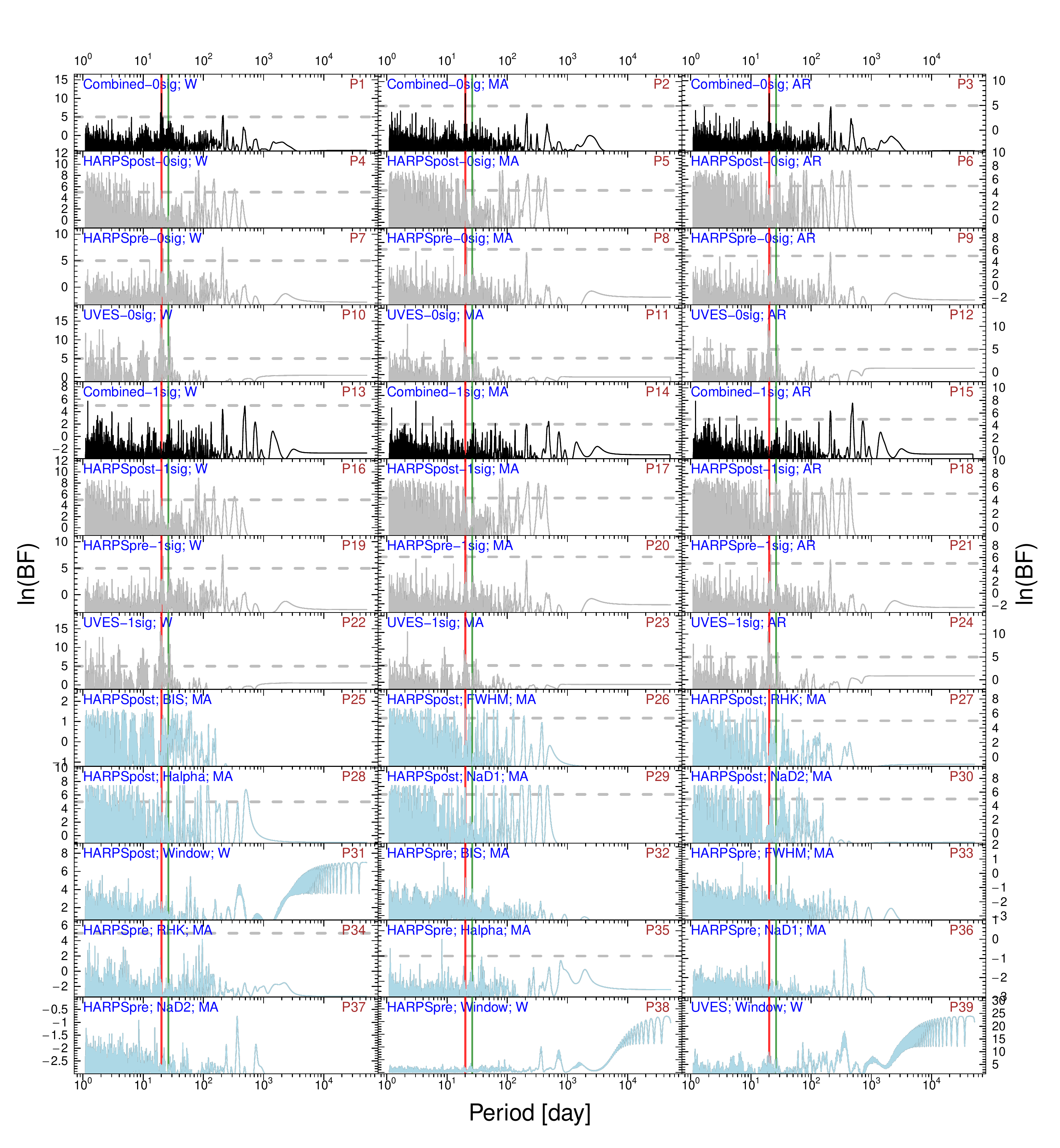}
  \caption{BFP for GJ 422. The red lines denote the signal at a period of
    20.1 days while the darkblue lines denote the
    activity signal with a period of 8.1 days. The green line
    denotes the 26-day signal found by T14. }
  \label{fig:BFP_GJ422}
\end{figure*}

\begin{figure*}
  \centering
  \includegraphics[scale=0.4]{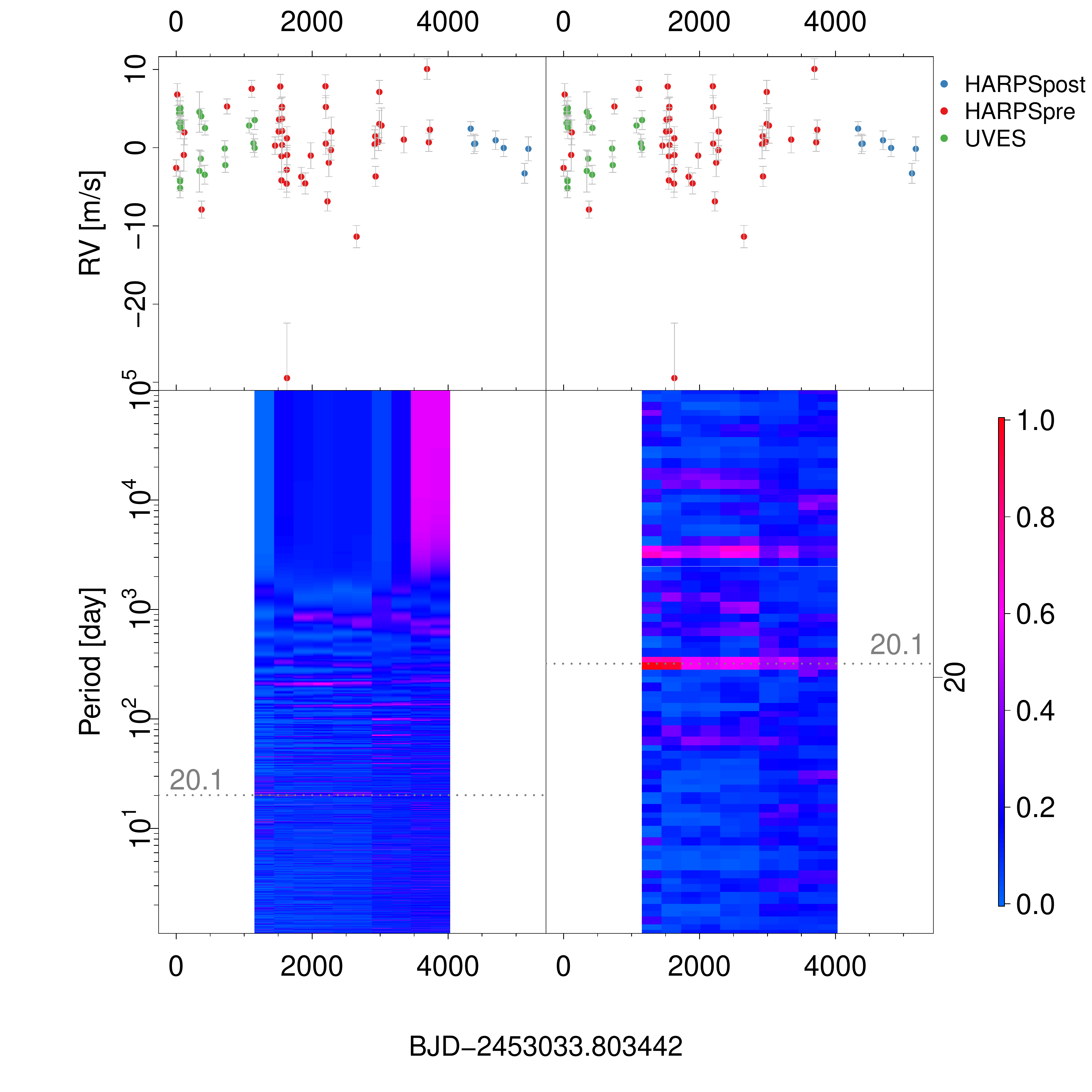}
   \caption{Moving periodogram for GJ 422 b. The annual aliases of the
     20.1-day signal at periods of 19 and 21 days are visible though not
     consistent over time. The 20.1-day signal is not significant
       in recent epochs as in earlier epochs due to low-cadence measurements
       in recent epochs.}
  \label{fig:MP_GJ422b}
\end{figure*}

\begin{figure*}
  \vspace{-0.3in}
  \hspace{-0.5in}\includegraphics[scale=0.5]{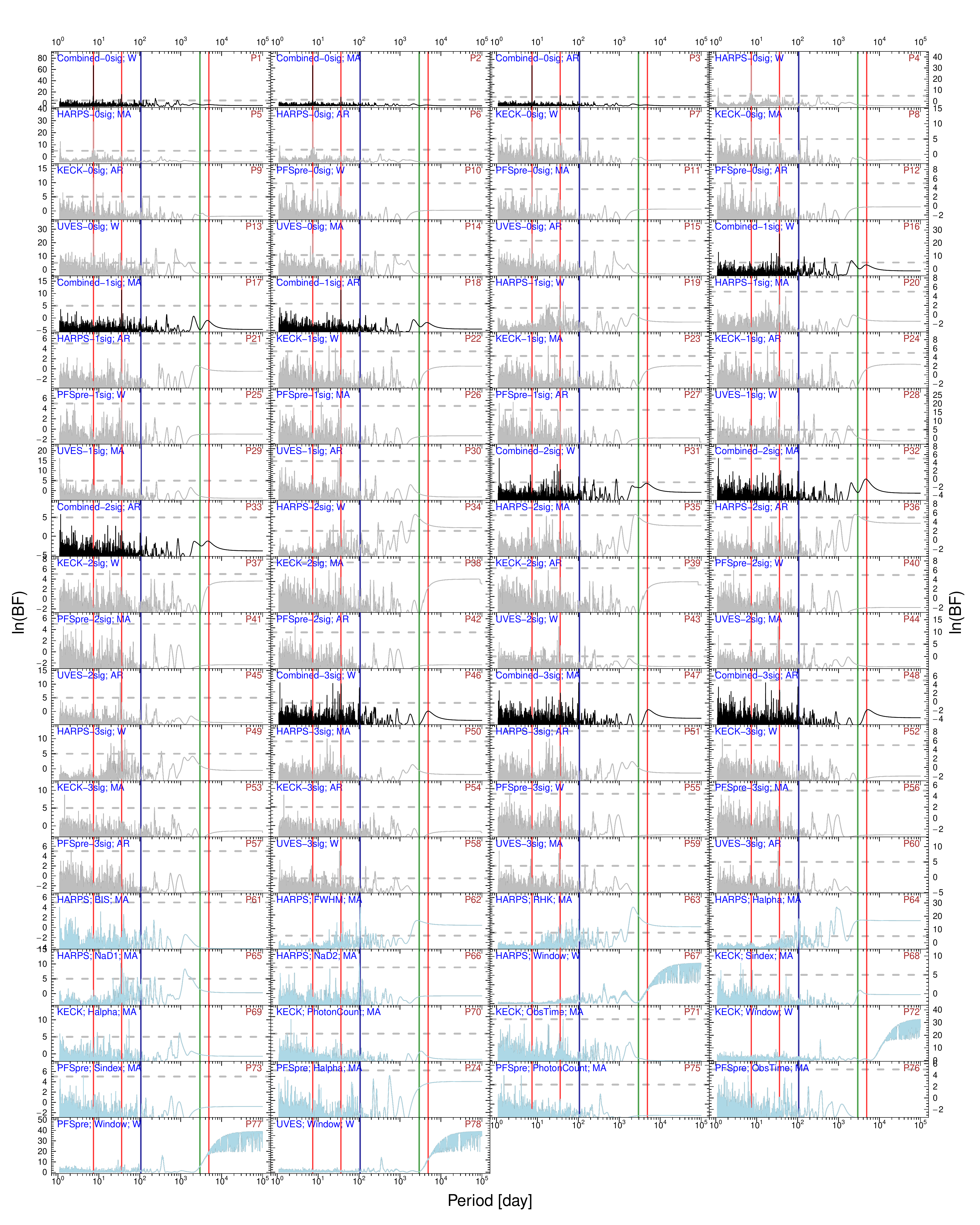}
  \vspace{-0.3in}
  \caption{BFPs for GJ 433. The red lines denote signal periods of
    7.37, 36.1 and 5090 days, which are subtracted from the raw RV data
    subsequently to calculate the residual BFPs. The darkblue lines denote the
    activity signal with a period of 107 days.}
  \label{fig:BFP_GJ433}
\end{figure*}

\begin{figure*}
  \centering
  \includegraphics[scale=0.4]{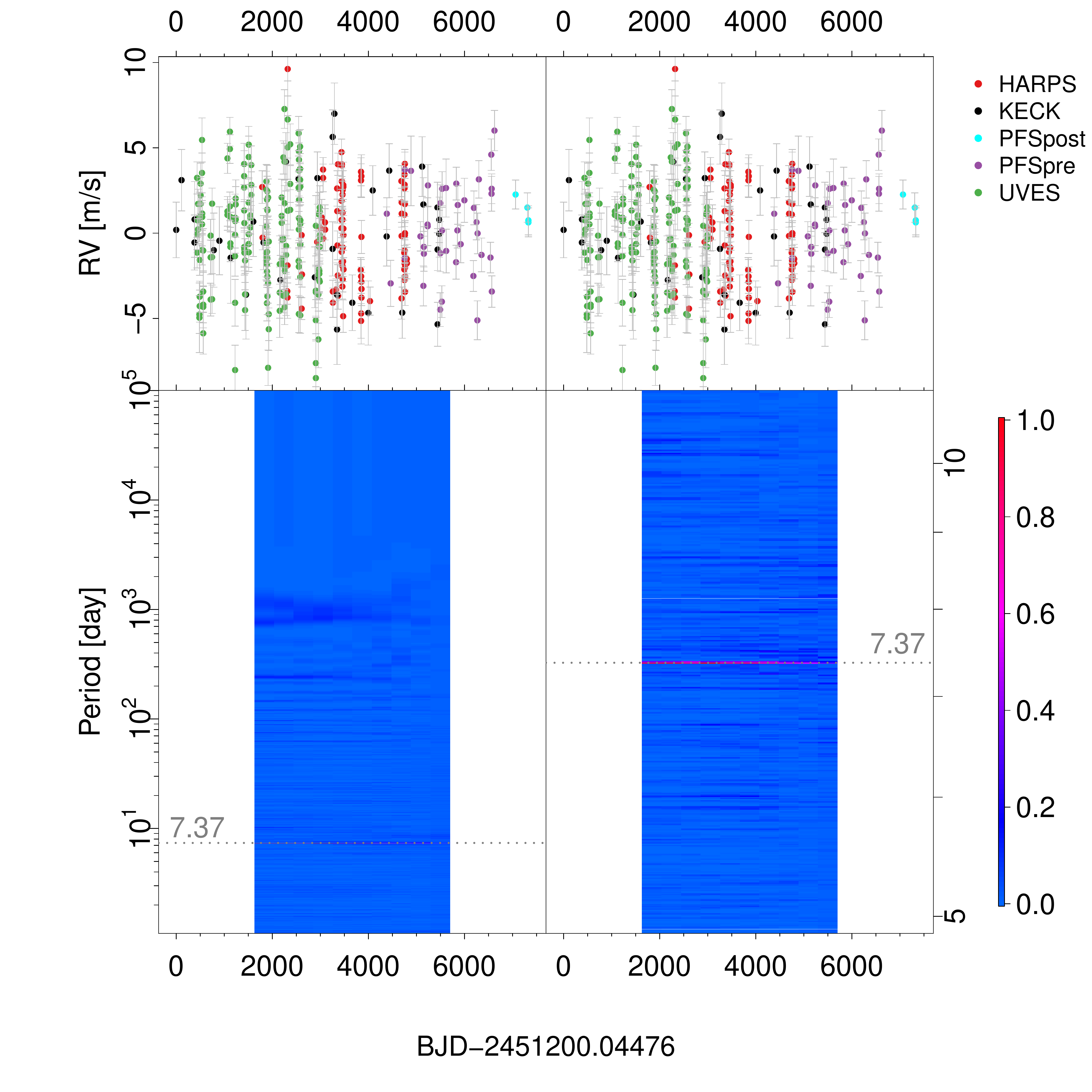}
   \caption{Moving periodogram for GJ 433 b. No annual aliases are
     visible. This signal is quite unique and consistent over time.}
  \label{fig:MP_GJ433b}
\end{figure*}

\begin{figure*}
  \centering
  \includegraphics[scale=0.4]{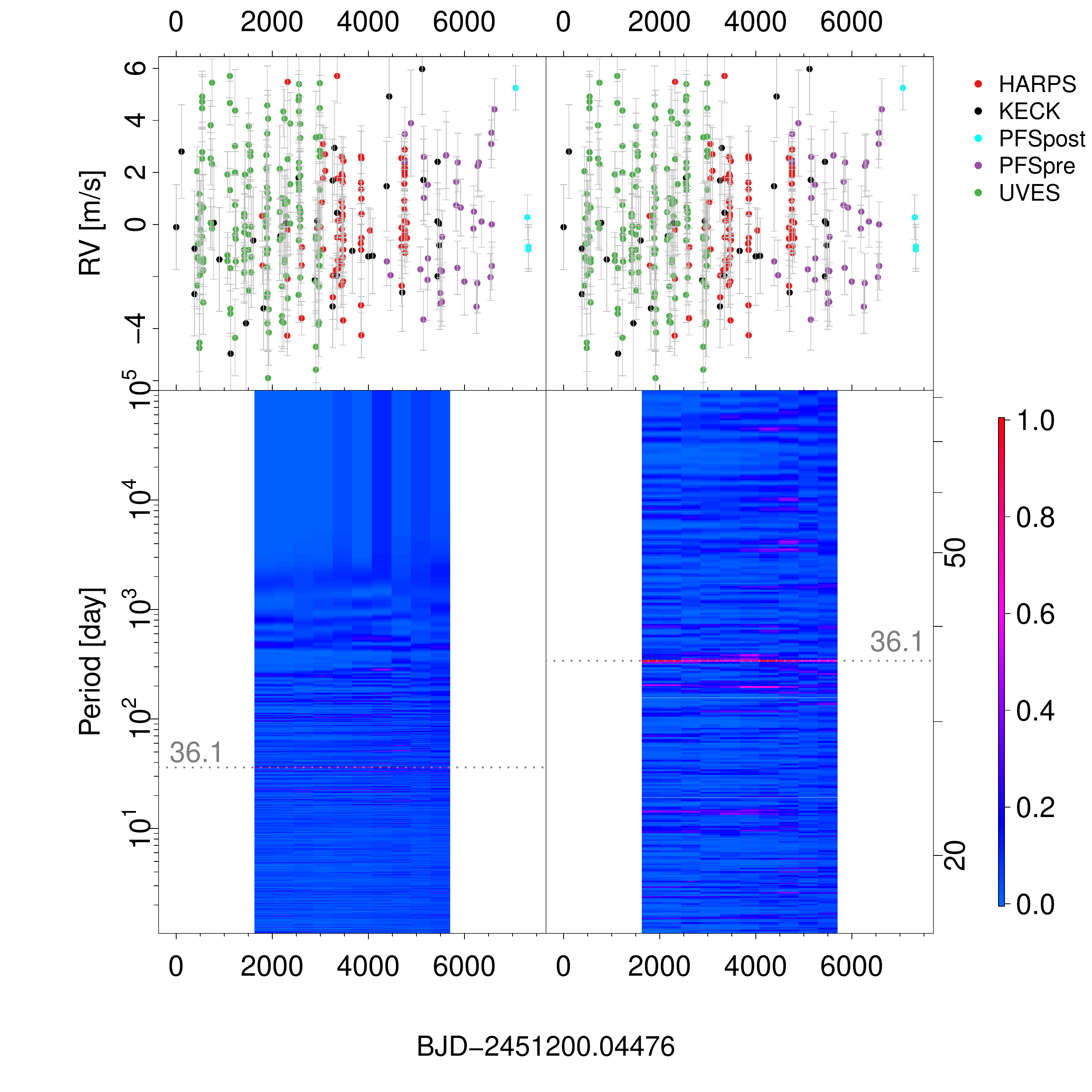}
   \caption{Moving periodogram for GJ 433 d. The annual aliases of the
     36.1-day signal at periods of 33 and 40 days are visible though not
     consistent over time.}
  \label{fig:MP_GJ433d}
\end{figure*}

\begin{figure*}
  \centering
  \includegraphics[scale=0.5]{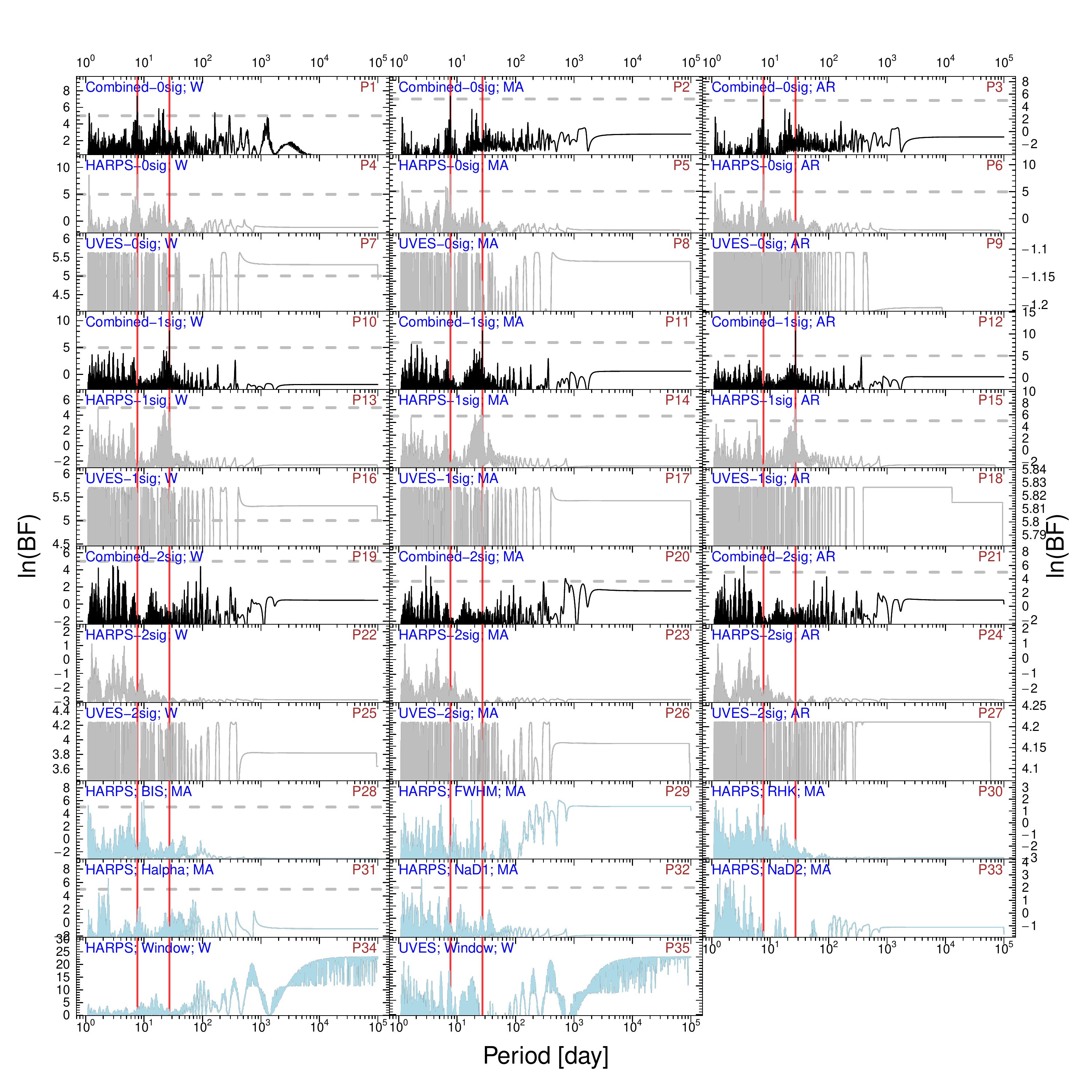}
  \caption{BFPs for GJ 620. The red lines show the signals at periods
    of 7.65 and 27.2 days, which are subtracted from the raw RV data
    subsequently to calculate the residual BFPs. }
  \label{fig:BFP_GJ620}
\end{figure*}

\begin{figure*}
  \centering
  \includegraphics[scale=0.4]{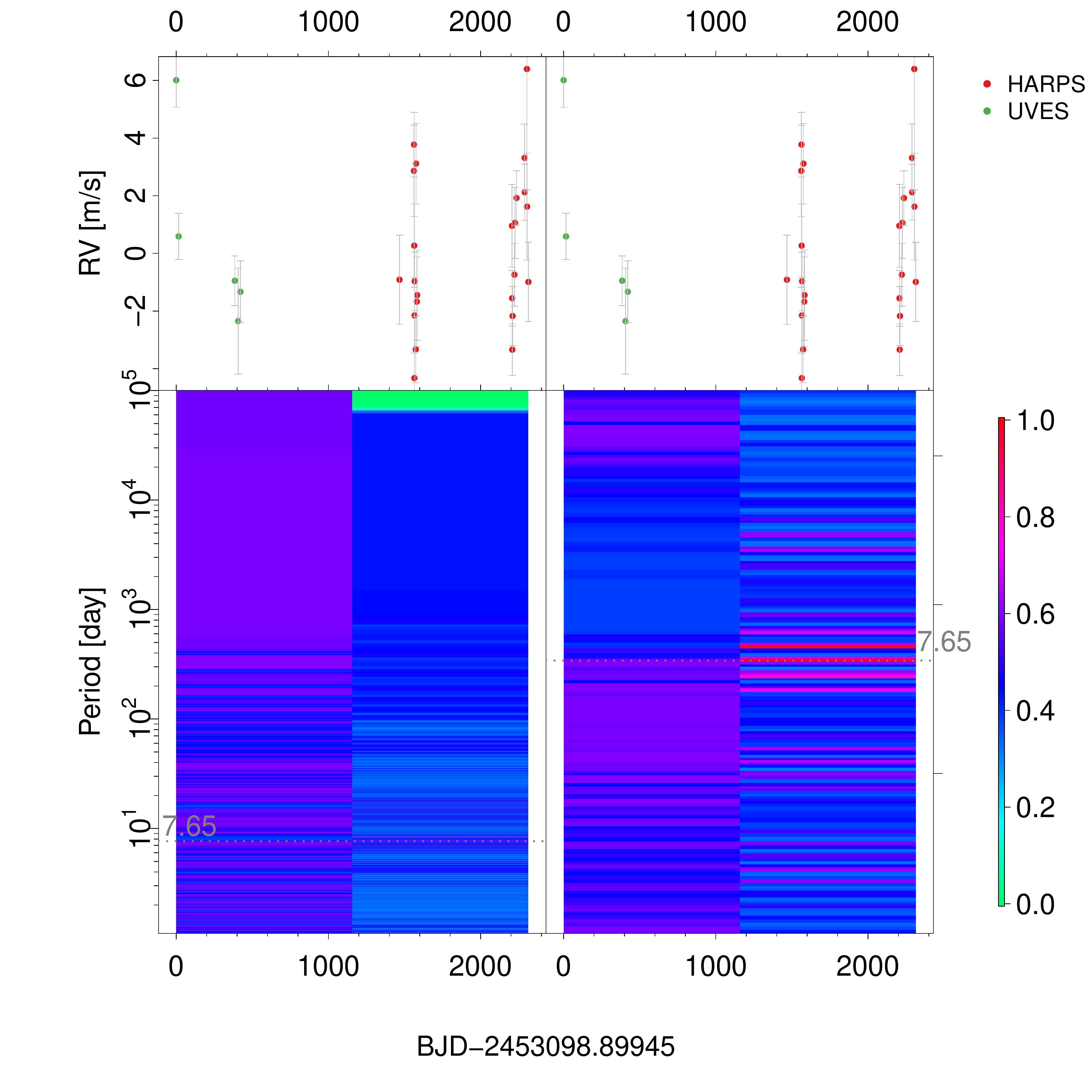}
   \caption{Moving periodogram for GJ 620 b. Considering the limited
     timespan, the combined data set are equally divided into two
     chunks. The signal at a period of 7.65 days is significant in the
     HARPS set and is consistent with the UVES set. }
  \label{fig:MP_GJ620b}
\end{figure*}

\begin{figure*}
  \centering
  \includegraphics[scale=0.4]{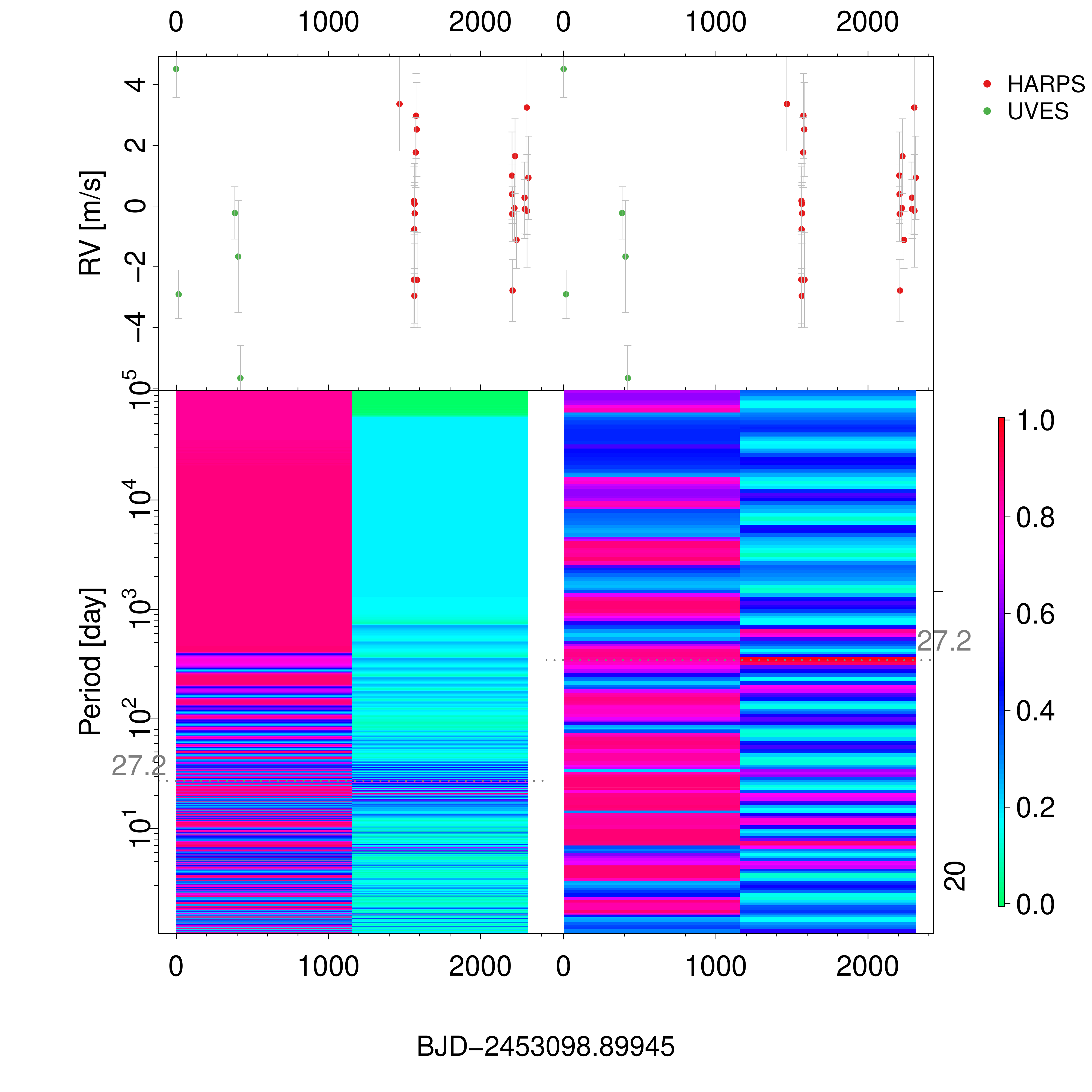}
   \caption{Moving periodogram for GJ 620 c. The signal at a period of
     27.2 days is significant in the
     HARPS set and is consistent with the UVES set. }
  \label{fig:MP_GJ620c}
\end{figure*}

\begin{figure*}
  \centering
  \includegraphics[scale=0.5]{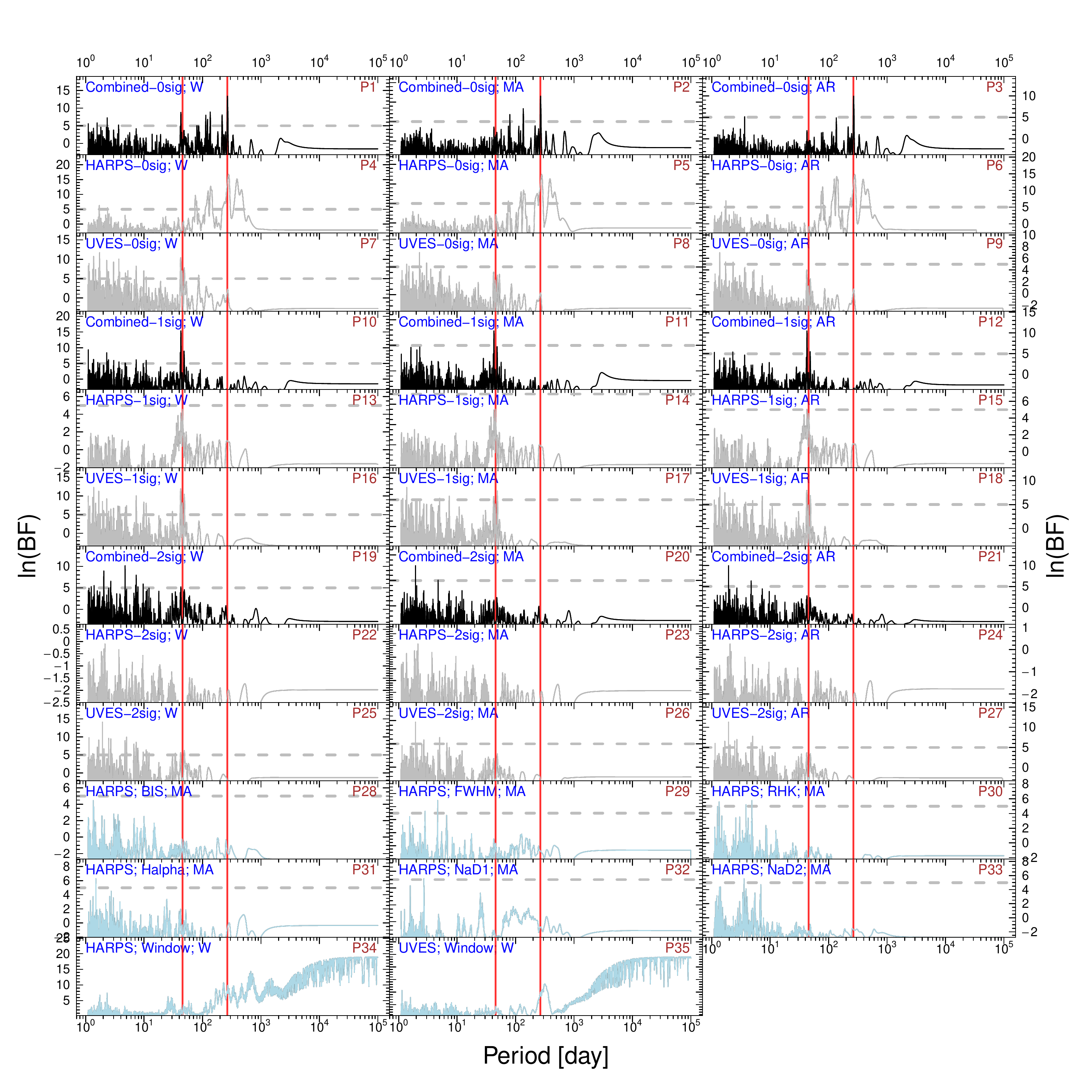}
  \caption{BFP for GJ 739. The red lines denote the signals at periods
  of 266 and 45.3 days, which are subtracted from the raw RV data
    subsequently to calculate the residual BFPs. }
  \label{fig:BFP_GJ739}
\end{figure*}

\begin{figure*}
  \centering
  \includegraphics[scale=0.4]{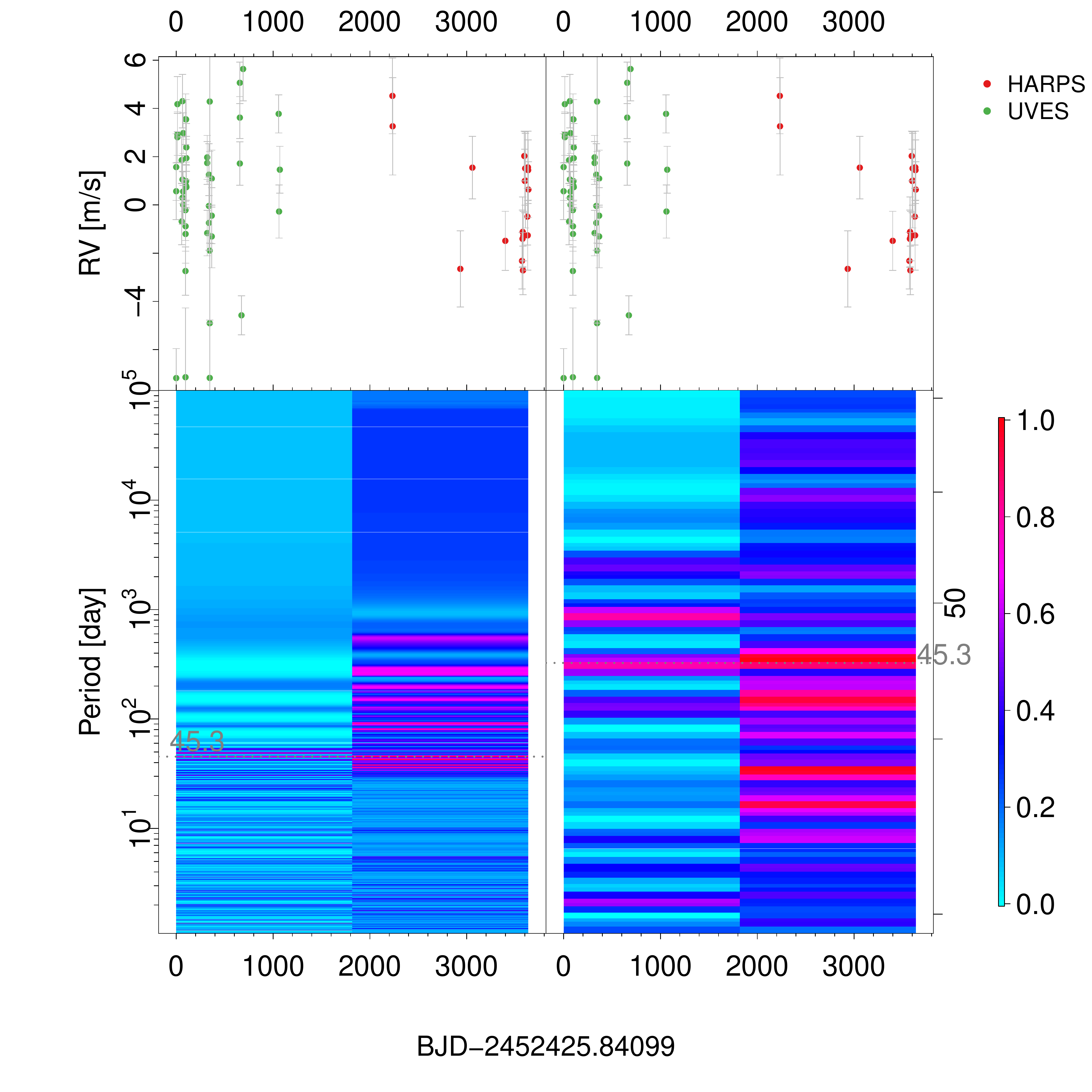}
   \caption{Moving periodogram for GJ 739 b. The signal at a period of
     45.3 days is significant in the
     UVES set and is consistent with the HARPS set.}
  \label{fig:MP_GJ739b}
\end{figure*}

\begin{figure*}
  \centering
  \includegraphics[scale=0.4]{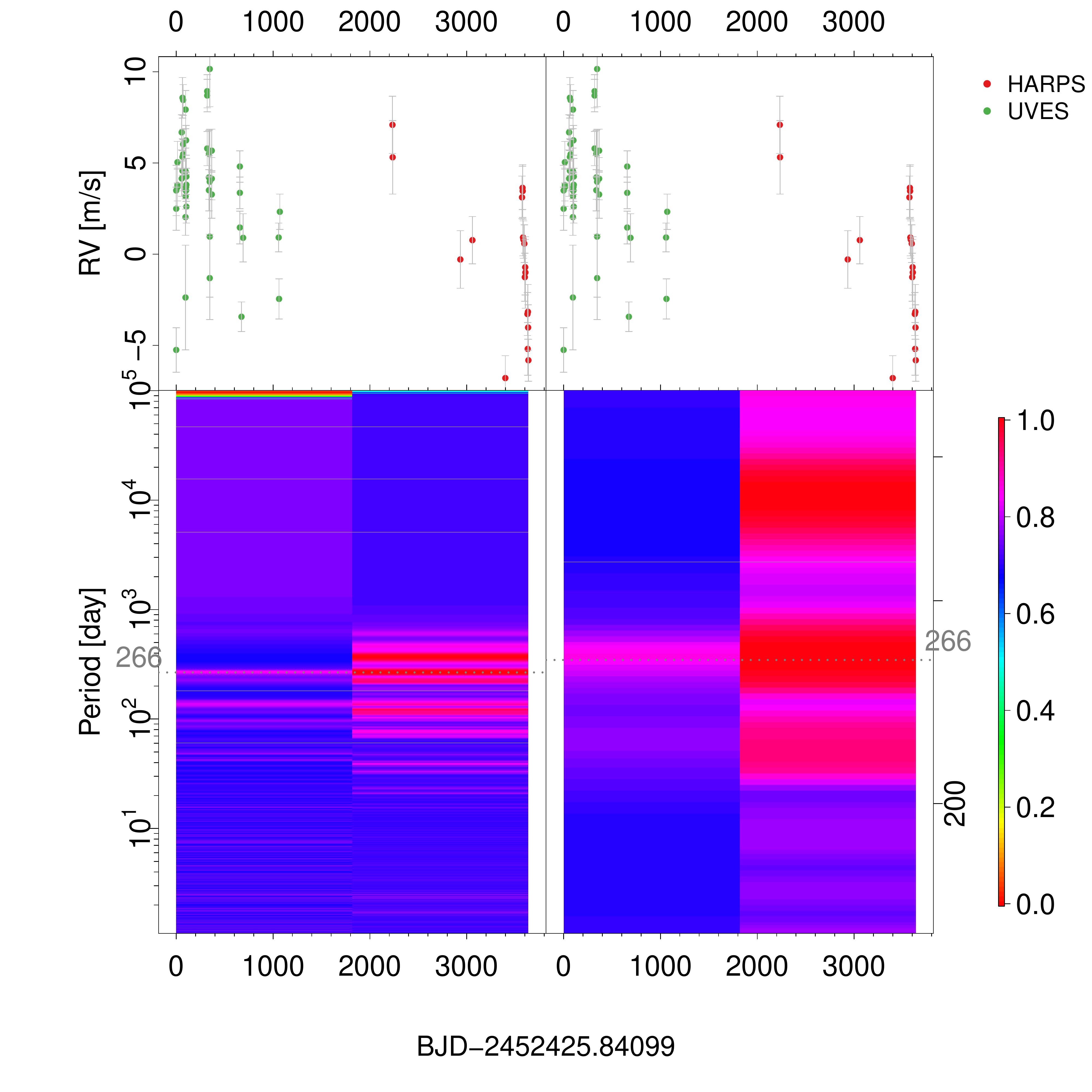}
   \caption{Moving periodogram for GJ 739 c. The 266-day 
     signal is strong both in the UVES and in the HARPS sets.}
  \label{fig:MP_GJ739c}
\end{figure*}

\begin{figure*}
  \centering
  \includegraphics[scale=0.5]{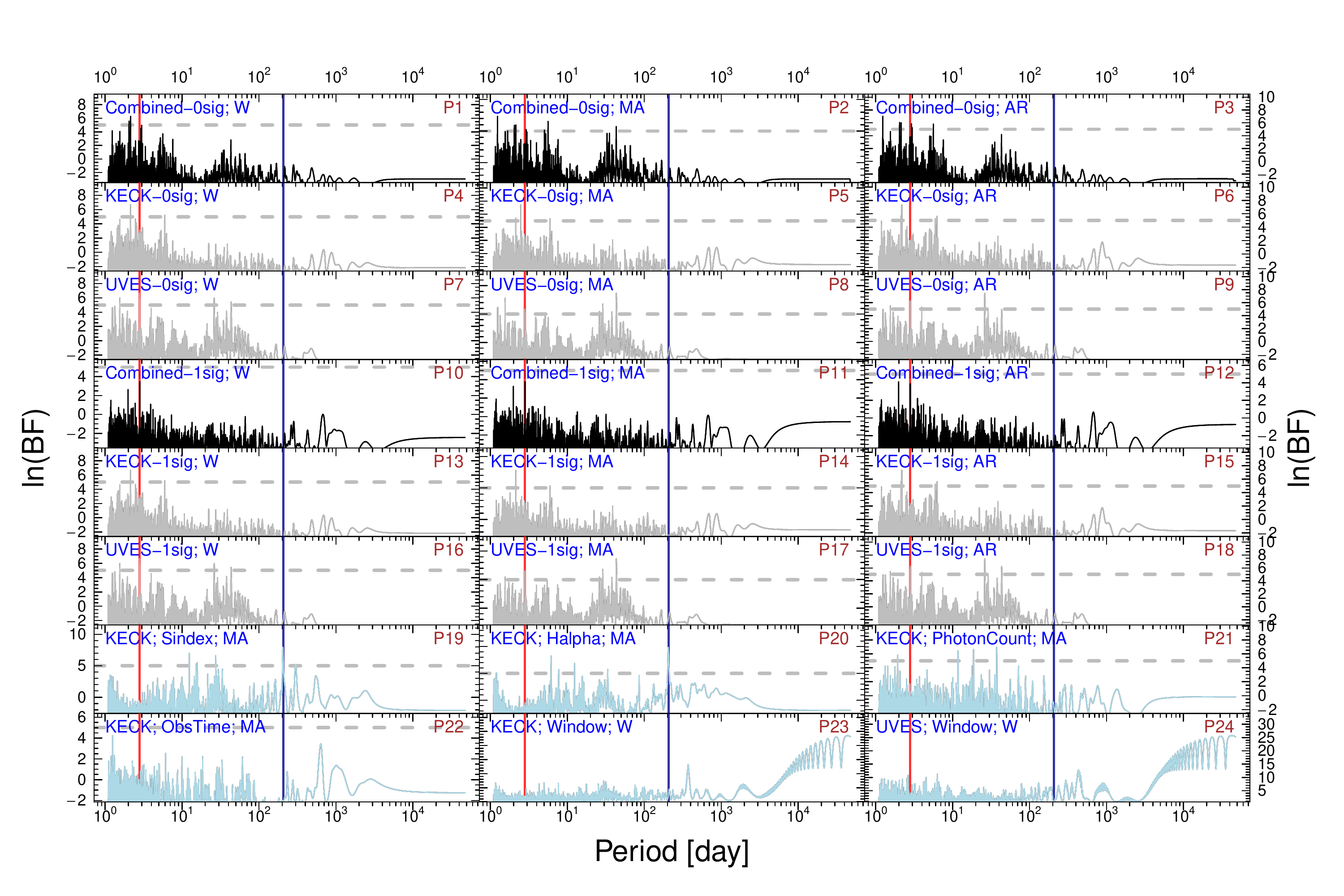}
  \caption{BFP for GJ 911. The red lines denote the planetary signal
    at a period of 2.79 days. The light darkblue lines denote the
    activity signal at a period of 204 days. }
  \label{fig:BFP_GJ911}
\end{figure*}

\begin{figure*}
  \centering
  \includegraphics[scale=0.4]{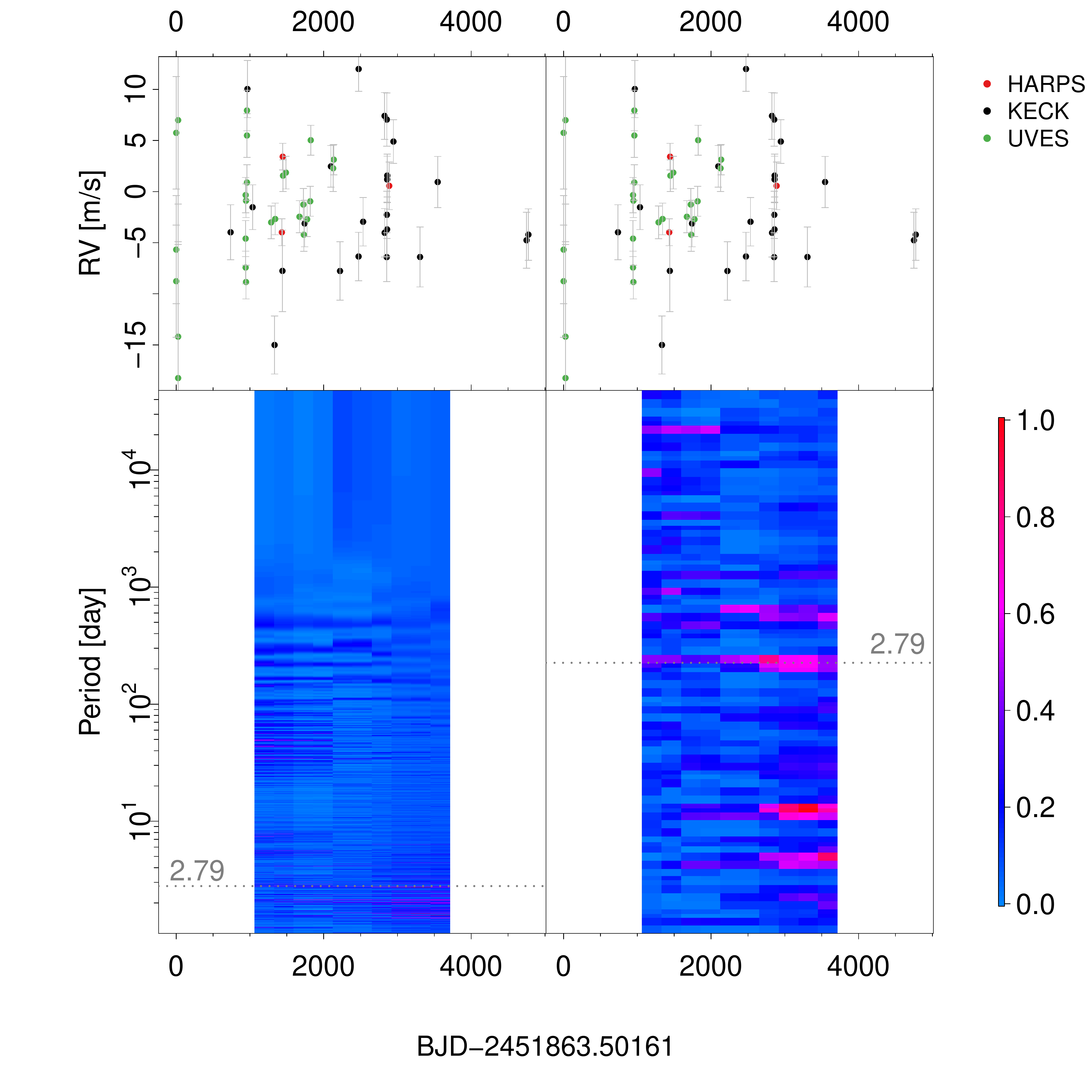}
  \caption{Moving periodogram for GJ 911. The annual aliases of the
    2.79-day signal at periods of 2.81 and 2.77 days are visible.}
  \label{fig:MP_GJ911b}
\end{figure*}

\begin{figure*}
  \centering
  \includegraphics[scale=0.5]{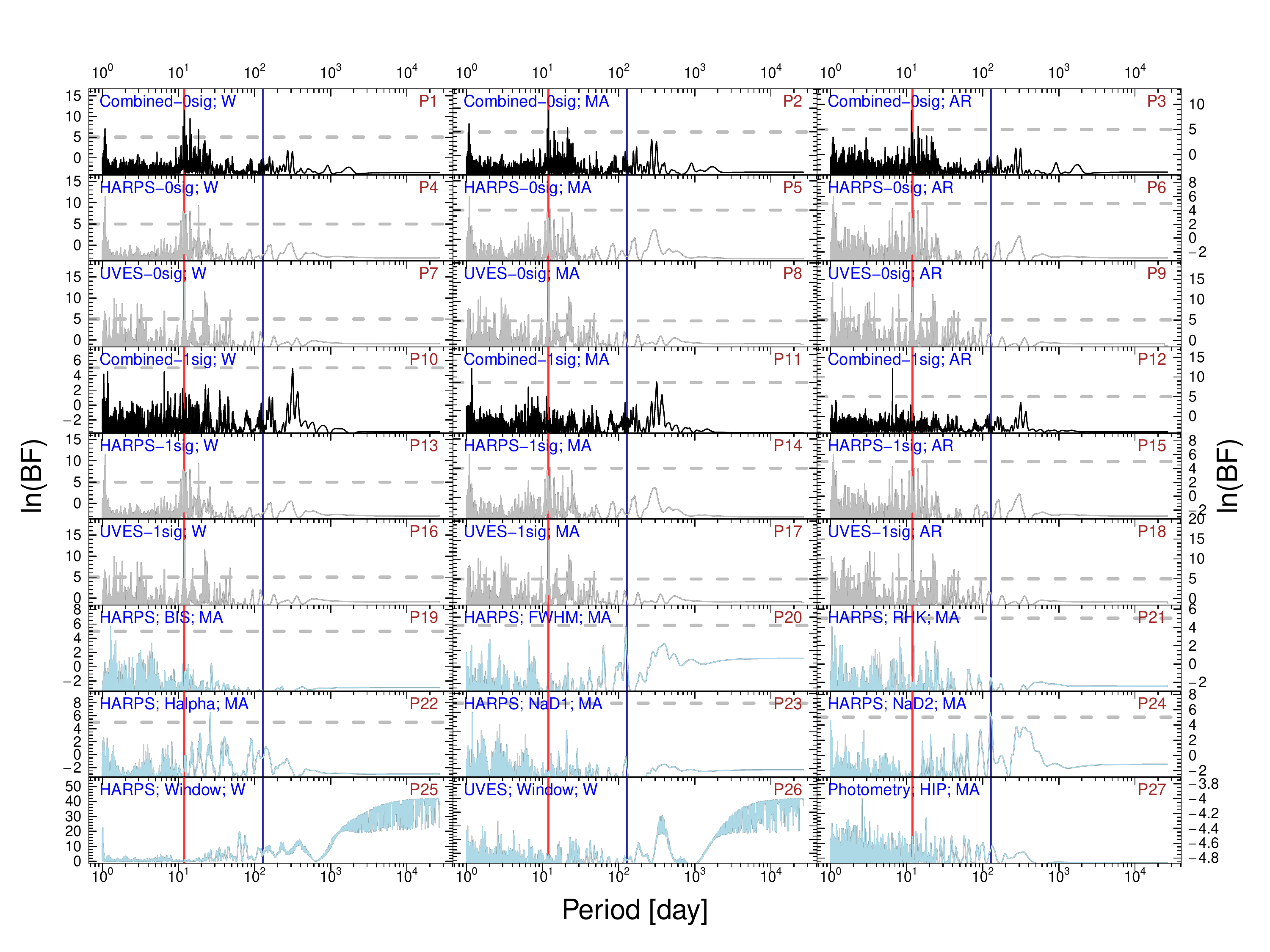}
  \caption{BFP for GJ 3082. The red lines denote the planetary signal
    at a period of 11.9 days. The darkblue lines denote the
    activity signal at a period of 123 days. The P27 panel shows the
    BFP for the Hipparcos photometry \citep{esa97}.}
  \label{fig:BFP_GJ3082}
\end{figure*}

\begin{figure*}
  \centering
  \includegraphics[scale=0.4]{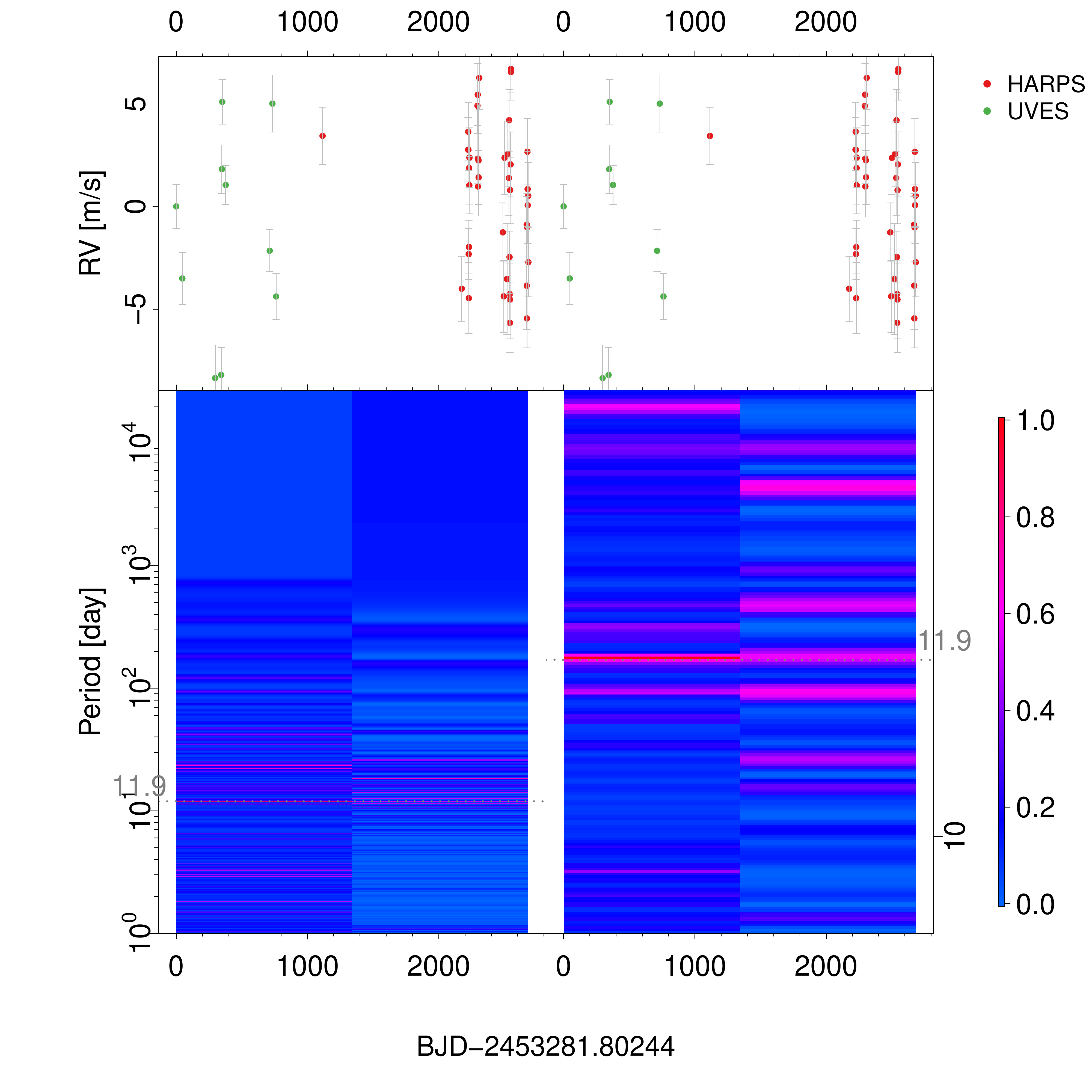}
   \caption{Moving periodogram for GJ 3082 b. The 11.9-day signal is
     consistently significant in HARPS and UVES sets.}
  \label{fig:MP_GJ3082b}
\end{figure*}

\begin{figure*}
  \centering
  \includegraphics[scale=0.5]{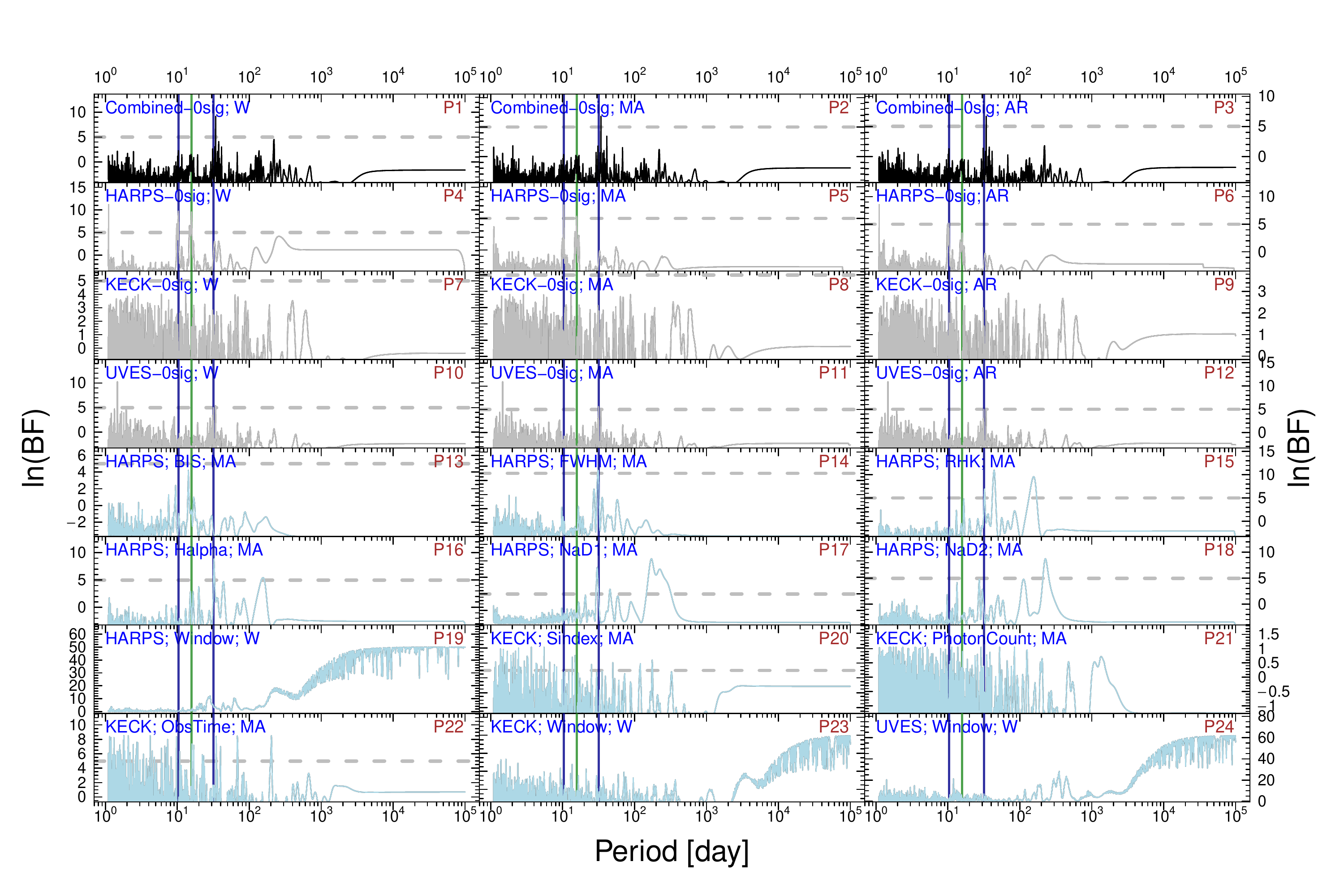}
  \caption{BFPs for GJ 27.1. The darkblue lines show the most
    significant activity signals at a period of about 31.8 and 10.3 days. The green line shows the signal at a period of 15.819 days identified by T14. }
  \label{fig:BFP_GJ27.1}
\end{figure*}

\begin{figure*}
  \centering
  \includegraphics[scale=0.5]{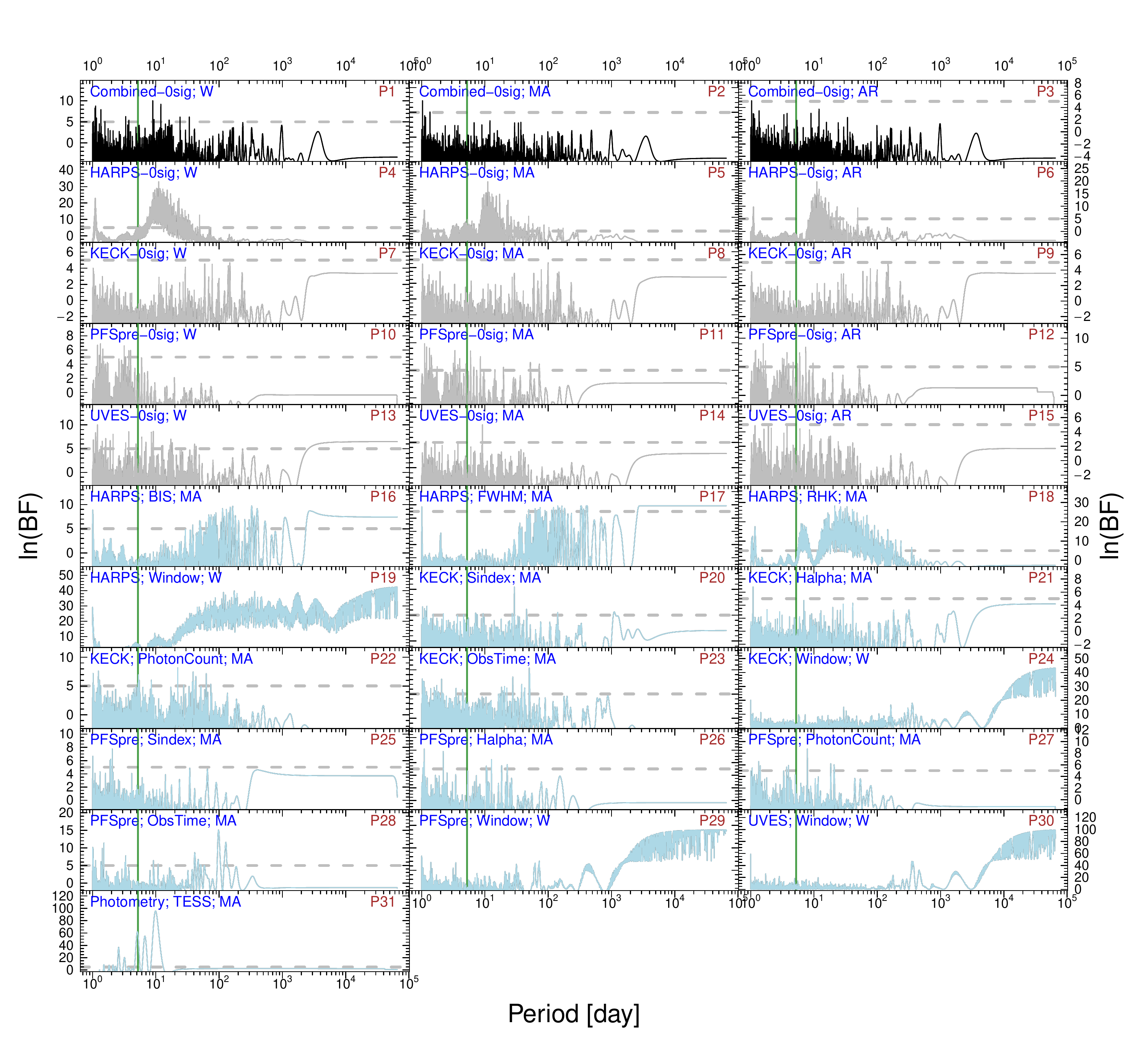}
  \caption{BFP for GJ 160.2. The green lines denote the signal at a
    period of 5.24 days found by T14. }
  \label{fig:BFP_GJ160.2}
\end{figure*}

\begin{figure*}
  \centering
  \includegraphics[scale=0.5]{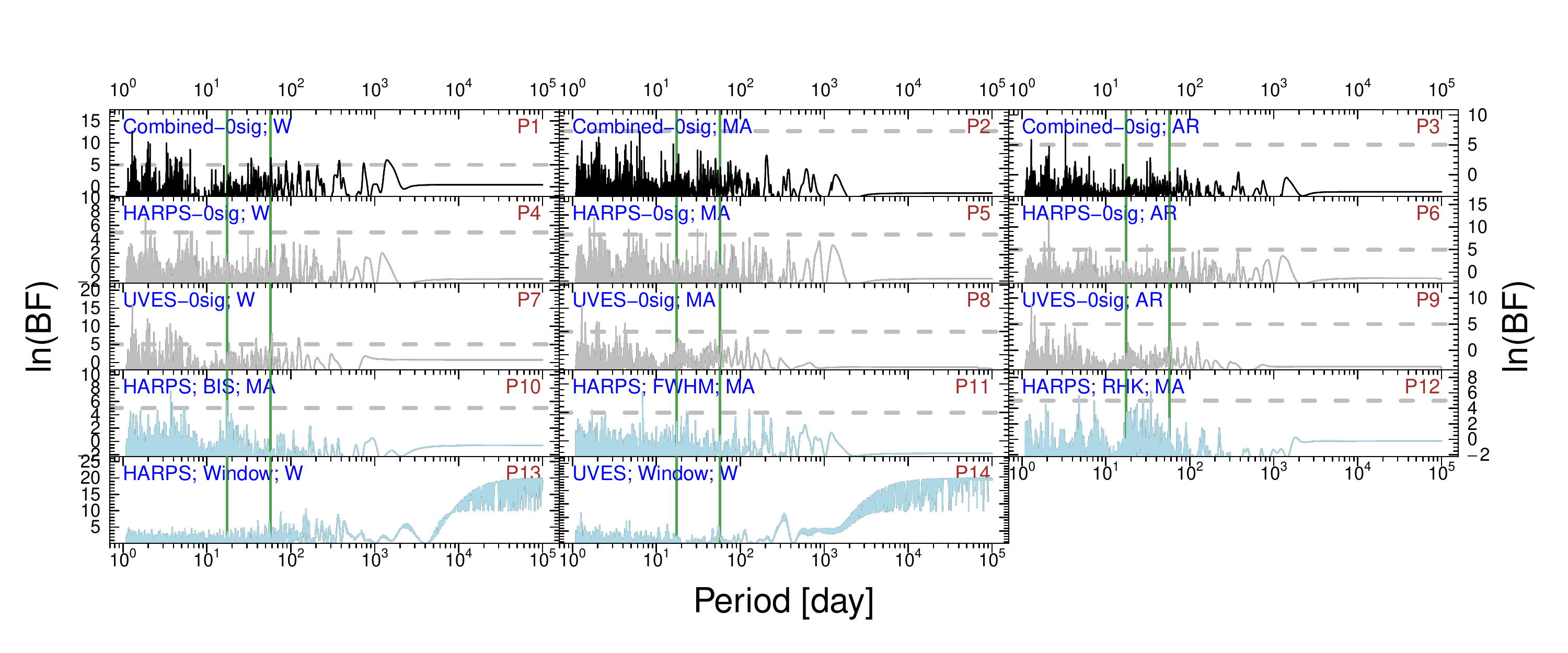}
  \caption{BFP for GJ 682. The darkblue line denotes the activity signal
    at a peirod of 6.87 days and the green lines denote the signals
    at 17.5 and 57.3 days reported by T14.}
  \label{fig:BFP_GJ682}
\end{figure*}